\def\simlt{\mathrel{\spose{\lower 3pt\hbox{$\mathchar''218$}}
     \raise 2.0pt\hbox{$\mathchar''13C$}}}
\def\simgt{\mathrel{\spose{\lower 3pt\hbox{$\mathchar''218$}}
     \raise 2.0pt\hbox{$\mathchar''13E$}}}
\documentstyle[aaspp4,epsf]{article}
\begin{document}
\def\gtorder{\mathrel{\raise.3ex\hbox{$>$}\mkern-14mu
             \lower0.6ex\hbox{$\sim$}}}
\def\ltorder{\mathrel{\raise.3ex\hbox{$<$}\mkern-14mu
             \lower0.6ex\hbox{$\sim$}}}
 
\def\today{\number\year\space \ifcase\month\or  January\or February\or
        March\or April\or May\or June\or July\or August\or
        September\or
        October\or November\or December\fi\space \number\day}
\def\fraction#1/#2{\leavevmode\kern.1em
 \raise.5ex\hbox{\the\scriptfont0 #1}\kern-.1em
 /\kern-.15em\lower.25ex\hbox{\the\scriptfont0 #2}}
\def\spose#1{\hbox to 0pt{#1\hss}}
\def\heion{\ion{He}{2}}
 
\title{Photometric Light Curves and Polarization of Close-in
Extrasolar Giant Planets}
\author{S. Seager\footnote{Institute for Advanced Study, Olden Lane, Princeton, NJ 08540}, B. A. Whitney\footnote{Space
Science Institute, 3100 Marine St., Suite A353, Boulder, CO 80303-1058},
D. D.~Sasselov\footnote{Astronomy Department, Harvard University,
60 Garden Street, Cambridge, MA 02138}$^,$\footnote{Alfred P. Sloan Research Fellow}}
 
\pagestyle{plain}

\begin{abstract}
The close-in extrasolar giant planets [CEGPs], $\ltorder$ 0.05 AU from
their
parent stars, may have a
large component of optically reflected light. We present theoretical
optical photometric light curves and 
polarization curves for the CEGP systems, from reflected planetary
light. Different particle sizes of
three condensates are considered. In the most reflective case, the
variability is $\approx 100$ micromagnitudes, which will be easily
detectable by the upcoming satellite missions MOST, COROT, and MONS,
and possibly from the ground in the near future. The least reflective case
is caused by
small, highly absorbing grains such as solid Fe, with variation of
much less than one micromagnitude. Polarization for all cases is 
lower than current detectability limits.
We also discuss the temperature-pressure
profiles and resulting emergent spectra of the CEGP atmospheres.
We discuss the observational results of $\tau$~Boo b by Cameron et al. (1999)
and Charbonneau et al. (1999) in context of our model results.
The predictions --- the shape and
magnitude of the light curves and polarization curves ---
are highly dependent on the size and type of condensates present in the 
planetary atmosphere.
\end{abstract}

\section{Introduction}
The discovery of the planet 51 Peg b in 1995 (Mayor \& Queloz 1995),
only 0.051 AU from its parent star, heralded an unexpected new class
of planets. 
Due to gravitational selection
effects, several more Jupiter-mass close-in extrasolar giant planets
(CEGPs) have been discovered since that time (Butler et al. 1997, 1998;
Mayor et al. 1999; Mazeh et al. 2000). To date there are 5 extrasolar
giant planets
$\ltorder$ 0.05 AU from their parent stars, and an additional 9 
$\ltorder$ 0.23 AU (see Schneider 2000). Relevant data about the close-in planet-star
systems (orbital distance $\ltorder$ 0.05 AU) are listed in
Table 1. Ongoing radial velocity searches 
will certainly uncover more CEGPs in the near
future. 
The CEGPs are being bombarded by radiation from their
parent stars, and could be very bright in the optical.
At best the CEGPs could be 
four to five orders of magnitude fainter than their primary star,
compared to Jupiter which is 10 orders of magnitude fainter than the
Sun.

The recent transit detection of HD~209458~b by Charbonneau et al. (2000)
and Henry et al. (2000b) confirms that the CEGPs are gas giants,
gives the planet radius, and fixes the orbital inclination, which
removes the $\sin i$ ambiguity in mass and provides the average planet
density. HD~209458 has $R_*=1.2 \pm 0.1 R_{\odot}$ and $R_P=1.40 \pm 0.17 R_J$ (Mazeh et al. 2000),
where $R_*$ is the stellar radius and $R_P$ is the planet
radius. Transits are definitely ruled out for $\tau$ Boo b, 51 Peg b,
$\upsilon$ And~b, HD~187123, and $\rho^1$ Cnc b,
whether they are assumed to be gas giants with radius 1.2 $R_J$, or
smaller rocky planets with radius $\sim$0.4 $R_J$
(Henry et al. 2000a, 1997; Baliunas et al. 1997; G. Henry, private communication).
Transits are also ruled out for HD~75289 (M. Mayor, private communication).
For a transit to be observable, a CEGP must be aligned with the star
as seen from Earth with an inclination $i > \theta_T$, where
$\theta_T={\cos}^{-1}((R_* + R_P)/D)$. For random orientations, the
probability for $i$ to be between $90^{\circ}$~and~$j^{\circ}$
is $P(j) = {\cos}(j)$. With $\theta_T \sim 83^{\circ}$, the CEGPs
have transit probabilities of ten percent. By the same $\theta_T$
criterion, the
non-detection of transits puts limits on the orbital inclinations to
approximately 83$^{\circ}$ (for $R_*=1.16$~$R_{\odot}$, $R_P=1.2$~$R_J$, and
$D=0.051$~AU). 
Several groups (e.g. STARE (PI T. Brown),
Vulcan Camera Project (PI W. Borucki), WASP
(PI S. Howell)) are monitoring
thousands of stars without known planets, searching with high
precision photometry for periodic
fluctuations indicative of a planetary transit.
Follow-up observations by radial velocity
techniques (or astrometry in the future) will be needed to fix the orbital radius in
order to determine the planet mass.
Edge-on CEGP systems are the most promising for reflected light signals.

Several observational approaches to detecting and characterizing
CEGP atmospheres have been developed. These include spectral separation,
transmission spectra observations during transit,
infrared observations, and optical photometric light curve observations.

Charbonneau et al. (1999) and Cameron et al. (1999)
have developed a direct detection technique: a spectral separation
technique to search for the reflected spectrum in the combined
star-planet light. Both groups have observed the 
$\tau$ Boo system. $\tau$ Boo A is one of the brightest (fourth
magnitude), hottest (F7V) parent stars, and
$\tau$~Boo~b has one of the smallest semi-major axes; these three
properties make $\tau$ Boo a promising candidate for this technique.
From a non-detection, Charbonneau et al. (1999)
have put upper
limits on the planet-star flux ratio ranging from $5 \times 10^{-5}$ for
$\sin i \sim 1$ to $1 \times 10^{-4}$ for $\sin i \sim 0.5$. The
result is within the strict assumptions
that the light curve is fairly isotropic and that the reflected
spectrum is an exact
copy of the stellar spectrum from 4668 to 4987 \AA.
Their upper limit on the geometric albedo
is 0.3 for $\sin i \sim 1$. 
The same technique for the $\tau$ Boo
system has been used by
Cameron et al. (1999) who claim a possible detection
at an inclination of 29$^{\circ}$, and give a planet-star flux ratio
of $1.9 \times10^{-4}$ at $i=90^{\circ}$. Given $R_P$, the albedo derived from this
type of observation can provide
a weak constraint on theoretical models.

A second approach is to observe
transmission spectra during a planet transit.
The stellar flux will pass through the optically thin
part of the planet atmosphere. Theoretical predictions
show the planetary absorption features will be
at the $10^{-4}$ to $10^{-3}$
level (Seager \& Sasselov 2000). Successful observations will constrain
the cloud depth and may give important spectral diagnostics such
as the presence of CH$_4$ which is a good temperature indicator
for the upper atmosphere layers.

A third technique under development is
the use of the Keck infrared interferometer in the differential phase
mode to directly detect and spectroscopically characterize
the CEGPs. The technique is based on the difference
between the very smooth infrared stellar spectrum
and the strong water absorption bands
and possibly methane bands in the CEGP's infrared spectrum.
See Akeson \& Swain (1999) for more details.

In this paper we present theoretical photometric light curves
and polarization curves of the CEGP systems.
As the planet orbits the star, the planet changes phase
as seen from Earth.
The planet and star are too close together for
their light to be separated, but this small separation means the stellar
flux hitting the planet is large, and the reflected light variation
in the combined light of the system from the planet's
different phases may be detectable. We focus on the optical
where there is a clear signature of reflected light: the planet's
dark side has no reflection or emission in
optical light. In contrast, there is no large light variation in the
infrared where the CEGPs are bright on
both the day and night side from reemission of absorbed
light. Scattered light is minimal in
the infrared, and is difficult to disentangle from the emitted
light. Unlike transits, which can only be seen for inclinations
$ > \theta_T$, the reflected light curves of lower inclinations are
theoretically visible and may be detectable.  
This work is motivated by upcoming microsatellites MOST ($\sim$2002) 
(Matthews 1997), COROT ($\sim$2003) (Baglin 1998, 2000), and MONS ($\sim$2003)
(Christensen-Dalsgaard 2000).
Initially intended for asteroseismology,
these satellites have capabilities to detect $\mu$mag
variability. MOST will observe known stars in a broad visual waveband, one
at a time for a period of roughly one month, including one with a
known CEGP in its first year. COROT's exoplanet
approach will use two CCDs in two colors to observe several fields of
$\sim$6000 stars for a few months each. Because of this wide-field
approach the stars with known CEGPs
will not be observed by COROT. While COROT's main exoplanet
focus is on transits, probability estimates suggest several CEGP
light curves from reflection should be detected. 
Precision of ground-based photometry on the CEGP parent stars
is currently at 100~$\mu$mag, and could reach 50~$\mu$mag
in the near future with dedicated automatic
photometric telescopes (Henry et al. 2000a). We also present
polarization signatures although they are well under the current
limits of detectability which is a few~$\times 10^{-4}$ in fractional
polarization of the system (e.g. Huovelin et al. 1989). 

This paper, to our knowledge, is the first to describe photometric
light curves and
polarization of CEGP systems: gas giants in close orbits around
Sun-like stars. Although our own Solar System
planets have been well studied in reflected and polarized light, the
CEGPs have effective temperatures an order of magnitude higher, so
completely different cloud species and atmospheric parameters are
expected. If observable, the light curves would roughly constrain
the type and size distribution of condensates in the
planetary atmosphere.
In \S2 we present definitions and analytical estimates of
reflected light from the CEGPs, in \S3 a description of our
model, and in \S4 results and discussion.

\section {Analytical Estimate of the Light Curves and Polarization}
\label{sec-analytical}
An analytical estimate of the amount of reflected light and
polarization of an EGP system is useful for both comparison with
simulations and for upper limit
predictions.
A good idealized case for such estimates is provided by modeling a
planet
as, for example, a Lambert sphere. The Lambert sphere derives from the
law of
diffuse reflection proposed by Lambert, postulating a reflecting
surface
with a reflection coefficient that
 is constant for all angles of
incidence.
The reflection coefficient is simply the ratio of the amount of light
diffusely reflected in all directions by an element of the surface to
the incident amount of light which falls on this element. The
general conditions postulated by Lambert,
for example angle independence, are satisfied strictly only
for an absolute blackbody
and an ideal reflecting surface (often called ``absolutely
white'',
``ideally matted'', etc.). Thus derives Lambert's definition of
albedo, with
its inherent ambiguities as discussed at the end of the 19$th$ century by
Seeliger, and the ultimate decision by Russell (1916) to endorse
Bond's (1861)
definition of albedo for use in the Solar System.
 
The arguments offered by Russell (1916) in favor of the Bond albedo
(over other albedo definitions in use at the time)
are 
still relevant for Solar System objects, but not necessarily for EGPs.
One important point in favor of the
Bond albedo was that it is derivable from observations.
The Bond albedo is defined as the ratio of the total amount of
reflected light to the total amount of	incident plane-parallel light
integrated over all angles.
Note that at the time of Russell --- before
multi-wavelength observations of the Solar System planets --- the Bond
albedo, $A$, was not defined as an integrated quantity over all wavelengths
as it is today. However, the discussion below is still valid with
either definition.
The Bond albedo $A$ can be separated into two quantities,
\begin{equation}
A = pq,
\end{equation}
where $p$ is the geometric albedo and $q$ is the phase integral.
The geometric albedo is defined as the
planet flux divided by the reflected flux from a perfectly diffusing
disk of the same radius.
The phase integral $q$ is defined as
\begin{equation}
q = \int_0^{\pi} \phi(\alpha) \sin\alpha d\alpha,
\end{equation}
where $\phi(\alpha)$ is the phase function, or the brightness variation
of the planet at different phases. 
The phase angle, $\alpha$,
is the angle between the star and Earth as seen from the planet;
$\alpha=0$
corresponds to ``opposition'' when the planet is maximally illuminated
as seen from Earth.
$p$ is measureable for all Solar System planets because it is a geometric
and photometric quantity. $q$ is measureable from Earth for Mercury,
Venus, Mars, and the Moon; for the outer planets whose phase angle 
variation is only up to several degrees from Earth satellite mission
observations were necessary.
Thus the benefit of the
Bond albedo: it is a physically meaningful quantity but
it can be determined empirically for the Solar System planets. 

In contrast, the Bond albedo cannot be determined from observations for 
extrasolar planets. Because the EGP systems are so distant from Earth,
only the CEGPs have prospects for measurement of $p$ in the
foreseeable future, and even then only the most reflective CEGPs 
will be bright enough,
and only $\sim$10\% of those will have orbital inclinations near $90^{\circ}$. Charbonneau et al. (1999) and Cameron et al. (1999) have developed
a spectral separation detection technique which can put upper limits
on --- and in the best case measure --- $p$ in a narrow wavelength region. 
For CEGPs that are reflective, and at $i\sim90^{\circ}$, $q$
should be measureable with the upcoming satellite missions.
However if a given
CEGP system is at $i<90^{\circ}$, the full range of phases will not be
visible (i.e. $\alpha$ will not be fully probed), and $p$ and $q$ will
not be measureable.
As discussed in \S\ref{sec-EGPbeyond}, EGPs beyond $D=0.1$~AU will not
be detectable in optically reflected light even with the upcoming
satellite
missions. The EGPs certainly have promise for detection in the infrared
where they emit most of their energy.
However, most of this energy is
reprocessed absorbed energy; it is not possible to measure
the Bond albedo with infrared observations.
To summarize, the Bond albedo came into standard usage because it
was a measureable
quantity. This is not possible for almost all of the EGPs because
of the distance of the systems and random orbital inclinations.

The goal of this paper is a presentation of
light curves at all viewing angles and at different inclinations, instead
of a Bond albedo.
We begin with the analytical estimate, where
in the idealized case
we are simply interested in the ratio, $\epsilon$, of the observed
flux
at Earth from the EGP at full phase ($\alpha=0$) to that of the star:
$\epsilon = p(R_P/D)^2$. Here $D$ is the star-planet distance,
and $p$ and $R_P$ are as previously defined.
For a Lambert sphere the single scattering albedo
$\tilde{\omega}=1$; no
photons are absorbed, and so $A=1$.
For a Lambert sphere, all incoming photons
are singly, isotropically scattered, and $p = 2/3$.
The light
variation of the Lambert sphere is only due to phase effects $-$ the
phase function, $\phi$, is simply (Russell 1916):
\begin{equation}
\phi(\alpha) = \frac{\sin(\alpha) + (\pi - \alpha)\cos(\alpha)}{\pi},
\end{equation}
and the phase-dependent flux ratio is ${\epsilon}\phi(\alpha)$.
Note that the phase angle, $\alpha$, is a function of the orbital
phase and inclination.
We convert this flux ratio to variation in micromagnitudes by
\begin{equation}
\label{eq:dm}
\Delta m = -2.5 {\rm log}_{10}(1 + \epsilon \phi(\alpha)) 10^6.
\end{equation}
 
\begin{figure}
{\plotfiddle{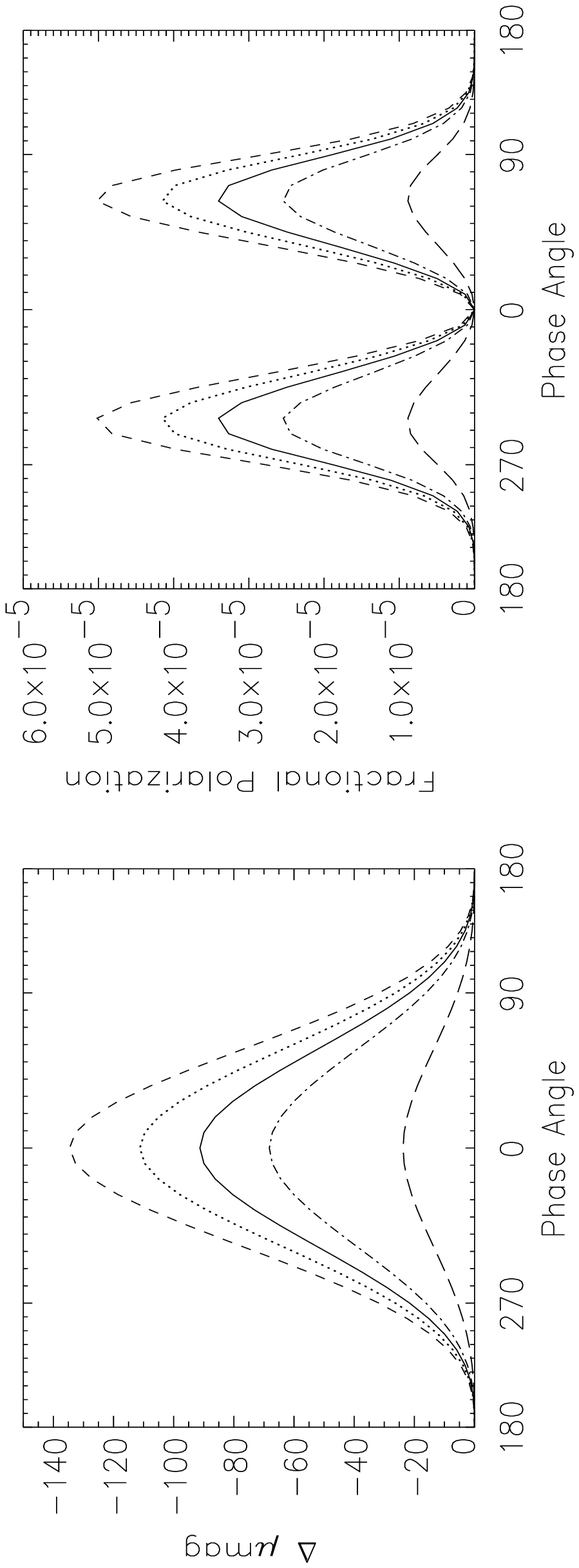}{2in}{-90}{80}{80}{-75}{310}}
\caption{Lambert sphere light curves and polarization curves for
CEGP systems with different $D$ and $R_P = 1.2 R_J$.
In descending order the curves are for $D=0.042$~AU
(HD 187123 b), $D=0.0462$~AU ($\tau$ Boo b), $D=0.051$~AU (51 Peg b),
$D=0.059$~AU ($\upsilon$ And b), and $D=0.11$~AU (55 Cnc b). For other
$R_P$ the curves
can be scaled --- the light curves approximately and
the polarization curves exactly ---
by the factor $(R_P/1.2 R_J)^2$. $\Delta m =2.5~\mu$mag corresponds
approximatley to a flux ratio of 10$^{-6}$ (see equation~(\ref{eq:dm})).}
\label{fig:lambert}
\end{figure}

Figure~\ref{fig:lambert}a shows the photometric light variation at
$i=90^{\circ}$ for a Lambert sphere of $R=1.2 R_J$ at various $D$ corresponding
to known CEGP systems. The large variation of
110 and 140 $\mu$mag for planets with $D$ corresponding to
$\tau$ Boo b and HD 187123 respectively is above current
ground-based limits (Henry et al. 2000a).
The Lambert sphere light curve is unrealistic, and
the point of this paper is to show that the situation
is far more complex and almost always conducive to a smaller light
variation for a few reasons.
Firstly, the single scattering albedo, $\tilde{\omega}$, is
generally
different from one. When optical photons are absorbed by condensates
or gas in the CEGP atmospheres, they are re-emitted in the infrared
or contribute to the thermal pool.
Secondly, multiple scattering gives more of a
chance for absorption over single scattering for the same single
scattering
albedo. Each time a photon scatters, its next encounter has a
scattering probability of $\tilde{\omega}$, but when a photon is
absorbed it
cannot contribute to the scattered light. This effect
depends on density, i.e. on the mean free path of the photon.
Thirdly, when particles are large
compared to the wavelength of light, the particle scatters
preferentially in the forward direction. In this case, photons that
enter the atmosphere are likely to be multiply scattered down into the
atmosphere and eventually absorbed,
rather than to be backscattered and escape the planetary
atmosphere.
 
Figure~\ref{fig:lambert}b shows the fractional
polarization of the total light of
a Lambert sphere at $i=90^{\circ}$ at various $D$ corresponding to
known CEGP systems, and assuming $R_P=1.2R_J$.
We assume Rayleigh scattering linear polarization
of unpolarized incident light $Pol_{Ray}= {{\sin^2} \theta_S}/({1
+ {\cos^2} \theta_S})$ (Chandrasekhar 1960). Here
$\theta_S$ is the scattering angle: $\theta_S=0^{\circ}$ is the forward
direction and
$\theta_S=180^{\circ}$ the backward direction. $Pol_{Ray}$ peaks at a
scattering
angle of $90^{\circ}$. For single scattering and $\tilde{\omega} = 1$,
all scattered light is polarized. The polarization signature plotted
in
Figure~\ref{fig:lambert}b does not peak at $\alpha=90^{\circ}$ because
the polarization is modulated by the reflected light curve; the
scattered light is maximally polarized at $\alpha=90^{\circ}$, but the
amount
of scattered light peaks at $\alpha=0^{\circ}$.
Plotted is $Pol = ({S_{\bot} - S_{\|}})/({S + F}) = \epsilon \phi(\alpha)
Pol_{Ray}$,
where $S$ is the total scattered light,
$F$ is the unpolarized stellar flux, $S \ll F$, and $S_{\bot}$ and
$S_{\|}$ are
the perpendicular and parallel components of the scattered light
respectively. In
reality polarization is much lower than this best case estimate, for a
few reasons. Firstly, the amount of scattered light is expected to be
lower as described above. Secondly,
for the case of multiple scattering not all of the scattered light is
polarized -- multiple scatterings mean the photon loses some of its
polarization signature. Thirdly, when the particle is large compared
to the wavelength
of light, different light paths through the particle and interference
effects cause the polarized light to be lower and to have more 
than one peak, and so a strong single peaked signal such as shown in
Figure~\ref{fig:lambert}b may not be reached.
 
As shown in Figures~\ref{fig:lambert}a~and~\ref{fig:lambert}b, the
light curves and polarization are very sensitive to $D$, since
$\epsilon \sim 1/D^2$. For example, the Lambert sphere at $D=0.042$~AU
(corresponding to HD~187123) has a light curve and
polarization curve with amplitude twice as high as a Lambert sphere at
$D=0.059$~AU (corresponding to $\upsilon$ And b). The light curve 
and polarization curve estimates are also sensitive to $R_P$. Both
can be scaled --- the light curve approximately and
the polarization curve exactly ---
for different $R_P$ by the factor $(R_P/1.2 R_J)^2$.
For HD 209458~b with $R_P=1.40 \pm 0.17 R_J$ (Mazeh et al. 2000), this factor
is 1.36.

\section{Model Atmosphere}
The model atmosphere code consistently solves for the planetary emergent
flux
and temperature-pressure structure by simultaneously solving
hydrostatic equilibrium, radiative and convective equilibrium, and
chemical equilibrium in a plane-parallel atmosphere, with 
upper boundary condition equal to the incoming radiation. The code is
described in Seager (1999) and is improved over our 
code described in Seager \& Sasselov (1998) in two major ways.
One is a Gibbs free energy minimization code to calculate 
equilibrium abundances of solids
and gases, the second is condensate opacities
for 3 solid species. So while in Seager \& Sasselov (1998) we considered
neither the depletion of TiO nor accurate formation of
MgSiO$_3$, in the new models we do.

We compute the photometric light curves and polarization curves with a
3D Monte Carlo
code (Whitney, Wolff, \& Clancy 1999), using the
atmospheric profiles (opacities and densities as a function
of radial depth) generated by the Seager
\& Sasselov code. While in
principle one could compute light curves from the model atmosphere
code described above, the Monte Carlo scheme can treat a much more
sophisticated scattering method, with anisotropic scattering and
polarization, than our model atmosphere program. Both are described
in this section.

\subsection{Chemical Equilibrum}
The equation of state is calculated using a Gibbs free energy
minimization method, originally developed by White, Johnson, \& Danzig (1958),
and followed up by a number of papers in the 1960s and 1970s including a
particularly useful one by Eriksson (1971). A detailed description
of this method can be found in those papers; more recent
treatments relevant for astrophysics are described in Sharp \& Huebner
(1990), Petaev \& Wood (1998),
and Burrows \& Sharp (1999). A more complete chemical equlibrium
calculation applied to Gliese 229 B is described in Fegley and Lodders
(1996).

We have selected the most important
species from Burrows and Sharp (1999) for brown dwarfs and from Allard
\& Hauschildt (1995) for cool stars. We used fits to the Gibbs
free energy from Sharp \& Huebner (1990), from Falkesgaard (private
communication), or fitted from the JANAF tables (Chase 1998) following the
normalization procedure described in Sharp \& Huebner (1990). 
We include ions using a charge conservation in place of
the usual mass balance constraint (equation (10) in Sharp \& Huebner 1990).
In the Gibbs method we include 27 elements, with 90 gaseous species
and 4 solid species:
H, He, O, C, Ne, N, Mg, Si, Fe, S, Ar, Al, Ca, Na, Ni, Cr, P, Mn, Cl,
K, Ti, Co, F, V, Li, Rb, Cs, CO, H$_2$, OH, SH, N$_2$, O$_2$, SiO, TiO, SiS,
H$_2$O, C$_2$, CH, CN, CS, SiC, NH, SiH, NO, SN, SiN, SO, S$_2$, C$_2$H, HCN,
C$_2$H$_2$, CH$_4$, AlH, AlOH, Al$_2$O, CaOH, MgH, MgOH, VO, VO$_2$,
CO$_2$, TiO$_2$, Si$_2$C,
SiO$_2$, FeO, FeS, NH$_2$, NH$_3$, CH$_2$, CH$_3$, H$_2$S, KOH, NaOH,
NaCl, NaF, KCl, KF, LiCl, LiF, CsCl, CsF, H$^+$,
H$^-$, H$_2^-$, H$_2^+$, Na$^-$, K$^-$, Li$^-$, Cs$^-$, Fe (solid),
CaTiO$_3$, Al$_2$O$_3$, MgSiO$_3$.

While a full treatment of all naturally occuring elements and $\sim$2500
compounds (e.g. Fegley \& Lodders 1996)
is possible, our
abridged choice is certainly adequate for a first prediction of
CEGP photometric light curves. There may be non-equilibrium chemistry
involved (e.g. photochemistry on our own Solar System planets)
that is not addressed by even a complete thermodynamical equilibrium
calculation.
In general, as the temperature decreases from the inner atmosphere to
the outer atmosphere, the metal gases are depleted into solids which
are efficient absorbers or reflectors. The three
condensate opacities we chose (solid Fe, MgSiO$_3$, and Al$_2$O$_3$) have very
different optical constants (see \S\ref{sec-opacities} and
\S\ref{sec-g}), and are among
the dominant solids expected at the relevant temperatures and pressures.

\subsection{Radiative Transfer}
The flux from the parent star travels through the planetary
atmosphere, interacting with absorbers and scatterers in a 
frequency-dependent manner. In a condensate-free CEGP atmosphere
(Seager \& Sasselov 1998),
blue light will Rayleigh scatter deep in 
the atmosphere where the density of scatterers is highest, while infrared
light will be absorbed high in the atmosphere due to strong absorbers
such as TiO and H$_2$O. Similar
results were found for dusty models by Marley et al. (1999),
although they considered an isolated planet of the equilibrium
effective temperature. In other words,
they assumed that the absorbed stellar flux
can be accounted for as thermalized intrinsic flux for the
calculation of the atmospheric structure.
For differences in the self-consistent treatment of irradiation 
and an isolated planet at the same $T_{\rm eff}$,
see Seager \& Sasselov (1998) and Seager (1999).

The equilibrium effective temperature is defined by
$T_{\rm eq}~=~ T_*(R_*/2D)^{1/2}[f(1-A)]^{1/4}$.
Here the subscript * refers to the parent star, $D$ is the star-planet
distance, $A$ is the Bond albedo, $f$=1 if the heat is evenly
distributed, and $f$=2 if only the heated side reradiates the energy.
Physically, $T_{\rm eq}$ is
the effective temperature attained by an isothermal planet
(after it has reached complete equilibrium with its star).

Our approach is to use $F_* = \sigma T_*^4R_*^2 /4 D^2$ as the incoming
flux and assume that $f$=1, as it will for planets with a thick
atmosphere, due to rapid zonal and meridional circulation patterns
(Guillot et al. 1996). The factor of 4 is due to the assumption that the
absorbed incoming radiation is efficiently distributed to all parts
of the planet: radiation incoming to a cross section of $\pi R_P^2$
is reemitted into $4 \pi R_P^2$.
With this approach we need not use $T_{\rm eq}$, nor the Bond albedo used
in the $T_{\rm eq}$ definition. Heating of the planet happens in a
frequency- and depth-dependent manner, and the heating, as well as
the planet's $T_{\rm eff}$, comes out of the model atmosphere solution.
We treat the incoming flux as plane-parallel, which is accurate
for isotropic scattering (see \S\ref{sec-size0.1}). 

An additional but small contribution to the $T_{\rm eff}$ is
the internal planetary heat.
Because of the strong irradiation, the
internal temperature of the planet is greater than the
internal temperature of an isolated planet (Guillot et al. 1996).
The planet possesses an intrinsic luminosity because it leaks some of
the heat acquired during formation by loss of gravitational energy.
Thus, the atmosphere's inner boundary
condition is age- and mass-dependent, and needs a self-consistent
atmospheric and evolutionary calculation with accurate irradiation
and spectral modeling.
In any case the reflected spectra are not affected by the lower
boundary condition; any good guess is too cool to produce light
in the optical, which is entirely reflected light.

We solve the radiative transfer equation using the Feautrier method with
100 angular points and 3500 wavelength points. We include isotropic
scattering except for Rayleigh scattering which can be added to
the Feautrier method via a modified source function (Chandrasekhar 1960).

\subsection{Opacities}
\label{sec-opacities}
We get the optical constants of MgSiO$_3$ (enstatite) from Dorschner et al. (1995),
of Al$_2$O$_3$ (corundum) from Koike et al. (1995) and Begemann et
al. (1997), and of Fe (iron) from Ordal (1985) and Johnson \& Christy
(1973). In all cases the optical constants were extrapolated below 0.2 $\mu$m.
The condensate opacities (absorption and scattering) were computed
using Mie theory for spherical particles with a version of the code
from Bohren \& Huffman (1983). The condensate opacities dominate
over gaseous Rayleigh scattering. 

For H$_2$O, which is the dominant infrared absorber, we use the
straight means opacities (Ludwig 1971). TiO is only present
very deep in the atmosphere for the hottest models; we use straight
means opacities from Collins \& Fay (1974). Even if the features do
not appear in the atmosphere, the opacity contributes to the
temperature-pressure structure. We also include
H$_2$-H$_2$ and H$_2$-He collision-induced opacities from Borysow et al.
(1997), and Rayleigh scattering by H$_2$ and He from Mathisen (1984). CH$_4$
opacities are taken from the GEISA database (Husson et al. 1994)
which is incomplete for the high temperatures of the CEGPs.
Unfortunately the only
existing optical CH$_4$ opacities are coefficients derived from
Jupiter (Karkoschka 1994), and we do not include them.

The alkali metals, noteably Na~I and K~I, are very important opacity sources
in brown dwarf spectra (Tsuji et al. 1999; Burrows, Marley, \& Sharp 2000).
The alkali metals' (Na, K, Li, Cs) oscillator strengths and energy levels
were taken from Radzig \& Smirnov (1985). We only include the
low-lying resonance lines which may have large absorption
troughs in the optical.
We compute line broadening using a Voigt profile with H$_2$ and He
broadening, and Doppler broadening.

Many better line lists exist, and other opacities that are present
in L dwarf spectra may also appear in the CEGPs, but they are not
necessary for a first
approximation of light curves. We plan to include them in
future work. While the opacities are necessary for a self-consistent solution of
the atmosphere profile, the
reflected spectra are not sensitive to small details in the infrared
spectra. 

\subsection{Condensates}
The atmospheric structure and emergent spectra of our ``dusty'' models are
highly dependent on condensates, as first noted for brown dwarf models in
Lunine et al. (1989), and subsequently by other modelers (e.g. Tsuji et
al. 1996). The CEGPs have an extra sensitivity to condensates
because the strong irradiation will heat up the upper atmosphere
according to the condensate amount and absorptivity (see
Figure~\ref{fig:TempProfile}). Also, because the
incoming radiation is strongly peaked in the optical, in contrast to
isolated brown dwarfs which have little optical emission, the condensates
will cause strong reflection or absorption in the optical.

We consider four different sizes of
condensates, based on the Solar System planets.
The cases are intended to
explore the expected size range, in part because the cloud
theories are limited and may be in enough error that such
assumptions are just as good. The particle sizes considered are mean
radius $\overline{r}=$
0.01~$\mu$m, 0.1~$\mu$m, 1~$\mu$m, and 10~$\mu m$. All have
gaussian size distributions with a standard deviation of 0.1 times mean particle
radius. This choice is narrow enough to attribute specific effects to
a given particle size, but wide enough to prevent interference
effects. Cloud particles in Venus have $r=0.85 - 1.15~\mu$m, and a
haze layer above that has particles with $r=0.2~\mu$m (Knibbe et
al. 1997). The clouds
on Jupiter range from an upper haze layer with $r=0.5~\mu$m, and lower
cloud decks with $r=0.75~\mu$m, and $r=0.45 - 50$~$\mu$m (Taylor \& Irwin
1999). We assume that particles are distributed homogenously
horizontally and vertically from the cloud base.
The limitations of this assumption are discussed in \S\ref{sec-CloudLayers}

For the light curve calculation with the Monte Carlo scattering code
we compute the scattering matrix elements (Van de Hulst 1957) from Mie
theory, which describe the anisotropic scattering phase function and
the polarization.

\subsection{Monte Carlo method for scattering}

We use the atmosphere structure generated in the Seager \& Sasselov
code and then compute the light curves and polarization curves using a
Monte Carlo scattering code. In principle it is possible to compute
the light curves with a model atmosphere code, but it is much more accurate
to use the Monte Carlo code 
since it can deal with anisotropic scattering, the spherical
geometry of the planet, and can
easily compute all viewing angles --- inclinations and phases ---
from one run. 

The basic principle of the Monte Carlo scattering method is that
photon paths and interactions are simulated by sampling randomly from
the various probability
distribution functions that determine the interaction lengths,
scattering angles, and absorption rates. 
Incoming photons at a given frequency travel into the atmosphere (to a
location sampled from a probability distribution function), and
scatter using random numbers to sample from probabilistic interaction
laws. At each scatter, the photon's polarization and direction changes
according to
the phase function. 
Photons are followed until they are absorbed (they can no longer
contribute to the reflected light), or until they exit the sphere. On
exit, the photons are binned into direction and location; the result is
flux and polarization as a function of phase and inclination.

The code we use was adapted from several previous codes
described in the literature (Whitney 1991; Whitney \& Hartmann 1992, 1993;
Code \& Whitney 1995; Whitney et al. 1999).
Improvements from previous versions include exact sampling
of the scattering phase function for any grain composition,
arbitrary atmospheric density profiles, and inclusion of
arbitrary opacity sources.
Phase functions of MgSiO$_3$,
Al$_2$O$_3$, and Fe are computed using Mie theory.
Additional opacities include
Rayleigh scattering by H$_2$ and He, and
absorption by H$_2$O, TiO, H$_2$--H$_2$, H$_2$--He, and H$^{-}$.
Once absorbed, the photons are considered destroyed
--- they contribute to
the thermal pool and no longer can contribute to scattered light.
The Monte Carlo code
uses the atmospheric structure (density profiles) and opacities computed 
from the detailed plane-parallel radiative and convective
equilibrium code of Seager \& Sasselov, and computes scattering
from a spherical planet with such an atmospheric structure.
(As long as the scale-height of the atmosphere is small
the plane-parallel approximation is sufficient to determine atmospheric structure).
At visual wavelengths, the contribution of thermal emission from the
planet is essentially zero and the reflected light can be treated
as a scattering problem, where the incident radiation comes from the nearby
star, and absorbed flux is ignored.
Because condensate scattering is coherent we follow only one wavelength at a time.
Because the CEGPs are so close to their parent stars that plane-parallel
irradiation may not be accurate, we use
the correct angular distribution of ${\tan}^{-1}(R_*/D)$ (see
\S\ref{sec-size0.1}).

The Monte Carlo scattering method is preferable over ``traditional''
radiative transfer techniques because it can treat complex geometries,
and its probabilistic nature gives
all viewing angles at once.
In the traditional plane-parallel method one can only solve along the
line of sight,
and must use the same angles for the incoming radiation as for the
outgoing radiation (i.e. the emergent spectra). For the plane-parallel
atmosphere models, one model must be computed for each phase angle and
for each
inclination. Our particular model atmosphere only considers isotropic
or Rayleigh scattering, but realistically anisotropic scattering is
important (see \S\ref{sec-size0.1}).
Polarization is complicated and unnecessary when
solving the model atmosphere for hydrostatic equilibrium,
and for radiative and convective equilibrium in the traditional method.

Because of the need to use very large numbers of photons
($10^7$ to $5 \times 10^8$) in order
to fully sample the probability distribution function space, the
Monte Carlo method cannot solve the model atmosphere problem (it is
slow for optically thick regimes),
although progress is being made in this direction for radiative
equilibrium but with only a few line opacities (Bjorkman \& Wood 2000).

\section{Results and Discussion}
As an example of a CEGP we use 51 Peg b (Mayor \& Queloz 1995).
There are many uncertainties about
the CEGPs, including mass, radius, gravity, composition, T$_{\rm
eff }$, etc. We have chosen
only one example out of a large range of parameter space:
$M = 0.47~M_{\rm J}$, log $g$ (cgs) = 3.2, metallicity that of the parent star,
and $R_P=1.2~R_J$. (Note that the CEGPs' radii depend on mass,
heavy element enrichment and parent star heating. Evolutionary
models show $R_P$ for
a given CEGP could
be larger than the known radius of
HD 209458b ($R_P=1.4~R_J$) or as small as 0.9$R_J$
(Guillot 1999)).
The incident flux of 51 Peg A (G2V, T$_{\rm eff } = 5750$,
metallicity
[Fe/H] = +0.21, and log $g$ (cgs) = 4.4 (Gonzalez 1998)) was calculated from
the model grids of Kurucz (1992).

\subsection{Atmospheric Structure and Emergent Spectra}
We leave the detailed discussion of spectra and irradiative
effects on temperature and pressure
profiles for a separate paper. However, 
because the condensate assumptions affect the temperature-pressure profile
and hence emergent spectra, we discuss the general properties here.
These models supersede those in Seager \& Sasselov (1998) since grain formation
and grain opacities are considered,
most noteably, TiO condenses out of the upper atmosphere.

\subsubsection{Theoretical Spectra: Main Characteristics}
\label{sec-BasicSpectra}
\begin{figure}
{\plotfiddle{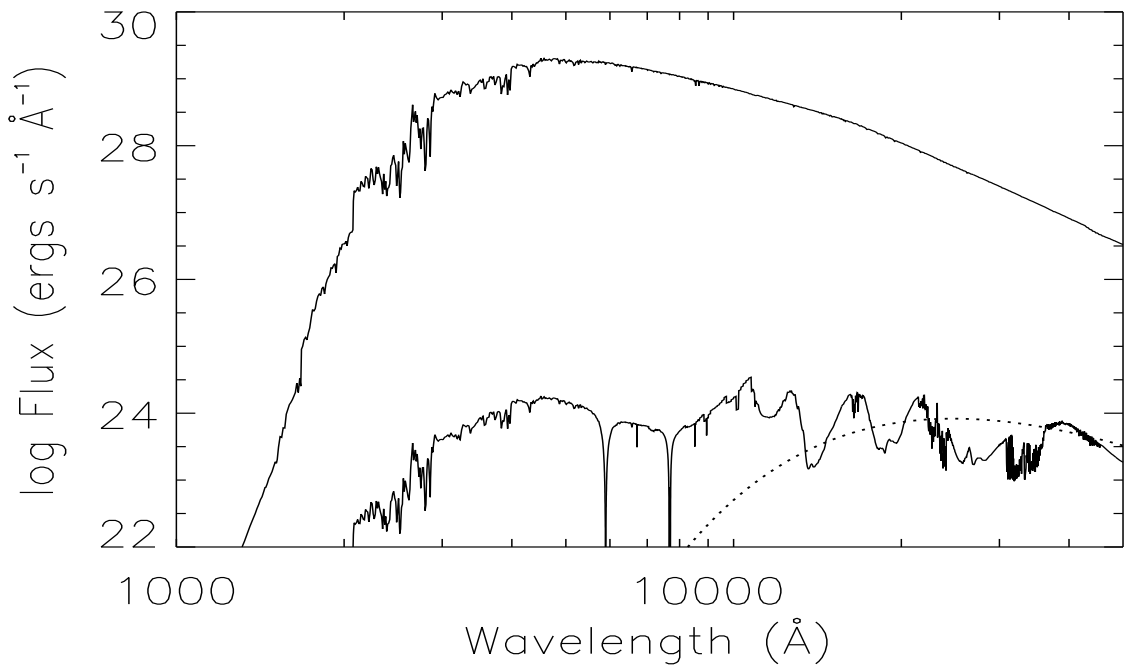}{3.5in}{0}{100}{80}{-85}{-300}}
\caption{Flux of 51 Peg A and b.
The upper curve is the model flux of 51 Peg A at the surface of the
star, and the lower curve is the model flux of 51 Peg b with a homogenous
MgSiO$_3$ cloud with particles of $\overline{r}=0.01~\mu$m. The dotted line is
a blackbody with the same $T_{\rm eff}$ as the planet model, 1170~K. In 51
Peg b, the features in the blue and UV ($<$ 5000 \AA) are reflected
stellar features, the absorption features between 5000 \AA~and
$1$~$\mu$m are alkali metal lines from the planetary atmosphere, and the
absorption features $> 1$~$\mu$m are water and methane absorption bands.}
\label{fig:BasicSpectra}
\end{figure}

Figure~\ref{fig:BasicSpectra} shows a model of 51~Peg~b with a
homogeneous cloud of MgSiO$_3$ particles with $\overline{r}=0.01$~$\mu$m,
with $T_{\rm eff}=1170$~K. The effective temperature $T_{\rm eff}$ refers
to the thermal emission only.
The most noticeable feature is the large optical flux, many orders of
magnitude greater than a blackbody ({\it dotted line}) of the same $T_{\rm
eff}$. The CEGPs have negligible optical emission of their own,
although the hottest ones at $T_{\rm eff}=1600$ K
may have some emission $>$ 7000 \AA, due to high absorption in the infrared
which forces flux blueward.
The CEGP in this model has 2--3 orders of magnitude more reflected flux
than an isolated planet
of the same $T_{\rm eff}$ has emitted flux. The CEGPs are at very
different effective temperatures from their parent stars of $\sim$6000~K, so
they have almost no molecular or atomic spectral features in common.
Thus the spectral features in
the blue and UV, blueward of $\sim$5200~\AA, are largely
spectral copies of the stellar spectra. Spectral features may also be
reflected at longer wavelengths where no absorbers are present,
for example H$\alpha$ at 6565 \AA.

The reflected optical component of the spectrum in this model comes
from reflection from a
homogeneous cloud of solid grains of MgSiO$_3$ with particles with
$\overline{r} = 0.01$~$\mu$m.
Rayleigh scattering from
H$_2$ and
He is negligible compared to the highly efficient scattering condensate,
but plays a role deep in the atmosphere.
The visual geometric albedo in this model is 0.18.
The reflected stellar features
between 4000 and 5200 \AA~ follow the slope of the scattering
coefficient of MgSiO$_3$.
In our models the condensate absorption and scattering features, such as
the well known 10~$\mu$m feature in comet reflectance spectra, do not
emerge in the CEGP spectra, since they occur where thermal emission of the
planet is strongest.

The absorption line at 7670 \AA~is the K~I $4^2p$ -- $4^2s$ resonance
doublet. Its broad wings
extend for several hundred \AA~and are responsible for the slope
redward to 1~$\mu$m. This effect 
is the cause of the large optical continuum depression in T dwarf spectra.
(Tsuji et al. 1999; Burrows et al. 1999).
This extreme broadening of the K~I resonance doublet is also seen in cool L
dwarf spectra (e.g. Tinney et al. 1999).
Such broad atomic absorption of this
kind --- wings of thousands of \AA~--- is not seen in any stellar
atmosphere and indeed came as quite a
surprise in the L dwarf observations. The cause is twofold: 1) strong pressure
broadening of a
fairly abundant species; and 2) there are no other strong absorbers in
that wavelength region.
The extreme broadening is not as surprising
if we consider, for example, that if the Sun had no other
absorbers than Lyman
$\alpha$ (at 1215~\AA) the wings would be visible out to the infrared
(R. Kurucz, private communication).
Other alkali metal lines are visible in the sample spectrum shown in
Figure~\ref{fig:BasicSpectra}: Na~I resonance doublet at 5893.6 \AA~($3^2p$ -- $3^2s$),
Cs~I at 8945.9 \AA~($6^2p_{1/2}$ -- $6^2s$) and
8523.5 \AA~($6^2p_{3/2}$ -- $6^2s$), Li~I at 6709.7 \AA~($4^2p$ -- $4^2s$)\
.
With very low ionization potentials --- between 3.89 and 5.39 eV --- the
alkali metals are in neutral atomic form for much of the temperature-pressure
regime in the CEGP atmosphere, although they do coexist with the
gaseous metal chlorides and fluorides in the very upper atmospheres. The
alkali metals are ionized in
stars, and form alkali metal chloride solids in cooler planets such as
Jupiter. Rb atomic lines should also be
present but are not included in our model atmosphere.
In principle Rayleigh scattering from Na~I and K~I could contribute a
small amount to the scattering (Dalgarno 1968), but would only become
important in condensate-free atmospheres.

The water bands are the most prominent absorption features in the
infrared, with broad absorption troughs at 1.15, 1.4, 1.9, and 2.7
$\mu$m. Because the depth of the troughs
is related to the temperature gradient, the spectral shape is
expected to change depending on the amount of upper atmosphere
heating, and to be different for irradiated planet atmospheres compared to
isolated planet atmospheres. The infrared flux is thermal
emission from absorbed and reradiated heat. Condensate absorption and
scattering affects this wavelength region as well, as described in the
next subsection.

The absorption trough at 3.3 $\mu$m is CH$_4$, and more minor methane
features are apparent at 1.6 and 2.3 $\mu$m.
The methane lines are
strong for this particular model. However as described in
the next subsection~(\ref{sec-condensates}) the presence of methane at all is very
sensitive to the amount of heating in the upper atmosphere.

\subsubsection{Effects of Condensates on Spectra}
\label{sec-condensates}
\begin{figure}
{\plotfiddle{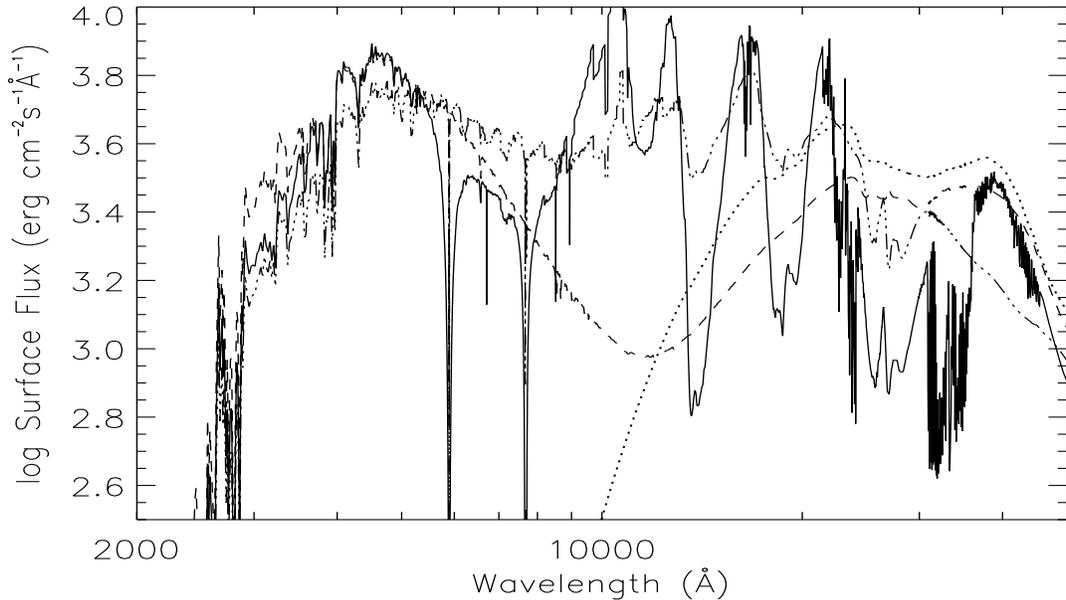}{1.5in}{0}{90}{70}{-85}{-275}}
\caption{Low resolution theoretical spectra of 51 Peg b
with homogeneous clouds of particles MgSiO$_3$-Al$_2$O$_3$-Fe in a
gaussian size distribution. The dotted line
is for particles with
$\overline{r}=0.01~\mu$m, dashed for $\overline{r}=0.1~\mu$m,
and dot-dash for $\overline{r}=10~\mu$m.
The solid line corresponds to the pure MgSiO$_3$ cloud with particles
with $\overline{r}=0.01~\mu$m, shown on a different scale
in Figure~\ref{fig:BasicSpectra}.
The features in the blue and optical are reflected stellar features
with the exception of the alkali metal lines. The
H$_2$O bands can be seen in the infrared. See text for details.}
\label{fig:spectra}
\end{figure}

The CEGP spectra are extremely dependent on the type, size, and amount of
condensates in the planetary atmosphere, and
Figure~\ref{fig:BasicSpectra} represents only one specific
model. Indeed it is impossible to predict the spectra,
albedo, or light curve without referring to a specific condensate mix and size
distribution. Figure~\ref{fig:spectra} compares different
low-resolution spectra at $\alpha=0$ of a subset of condensate
cases considered in this paper. The curves are spectra from the following condensate
assumptions: 
the solid line is a model with MgSiO$_3$ particles of mean radius
$\overline{r}=0.01~\mu$m (shown in Figure~\ref{fig:BasicSpectra}),
the dotted line is a model with a MgSiO$_3$-Fe-Al$_2$O$_3$ mix of particles
of $\overline{r}=0.01~\mu$m, the dashed line is a model with a
MgSiO$_3$-Fe-Al$_2$O$_3$ mix of particles
of $\overline{r}=0.1~\mu$m, and the dot-dashed line is a model with a
MgSiO$_3$-Fe-Al$_2$O$_3$ mix of particles
of $\overline{r}=10~\mu$m.

The dotted curve is the most absorptive case which resembles a
blackbody of $T_{\rm eff}$ close to $T_{\rm eq}$.
Reflected features (not visible) appear at a very low magnitude.
The most noticeable feature in the dashed curve is the broad
dip between approximately 3000--10,000~\AA. The cause is Rayleigh
scattering from the condensates, which in this wavelength region have
$\overline{r} \ll \lambda$. The slope is
$\sim \lambda^{-4}$, but displaced compared to gaseous Rayleigh scattering since the
Rayleigh scattering criterion is valid in a region of longer
wavelength. The reflected spectral features are still visible on this
Rayleigh scattering slope, in between the planetary atomic absorption
lines. The K~I and Na~I lines are visible, but the extreme broadening shown in
the solid line is not present because
scattering and absorption high in the atmosphere means the deep
atmosphere where the pressure broadening occurs is not sampled. 

The most noticeable difference in the dot-dashed curve compared
to the other spectra in Figure~\ref{fig:spectra}
is the presence of weak TiO features in the optical.
Condensates with $\overline{r}=10$~$\mu$m have more
scattering relative to absorption, and less absorption overall, right across the
wavelength range. In this case, incoming radiation
penetrates deep into the atmosphere, heating the atmosphere
over a large depth to temperatures where enough TiO
is present to produce weak absorption features.

CH$_4$ is an excellent temperature diagnostic for the CEGPs' upper
atmospheres.
The strong CH$_4$ 3.3 $\mu$m band, and weaker CH$_4$
features at 1.6 and 2.3 $\mu$m are present only in the coolest, least
absorptive model (solid line). The H$_2$O bands also differ
among the different models. They are much shallower
for the atmospheres with absorptive
condensates, due to absorption of incoming light by the condensates. 

\subsubsection{Temperature-Pressure Profiles}
\label{sec-TPprofile}
\begin{figure}
{\plotfiddle{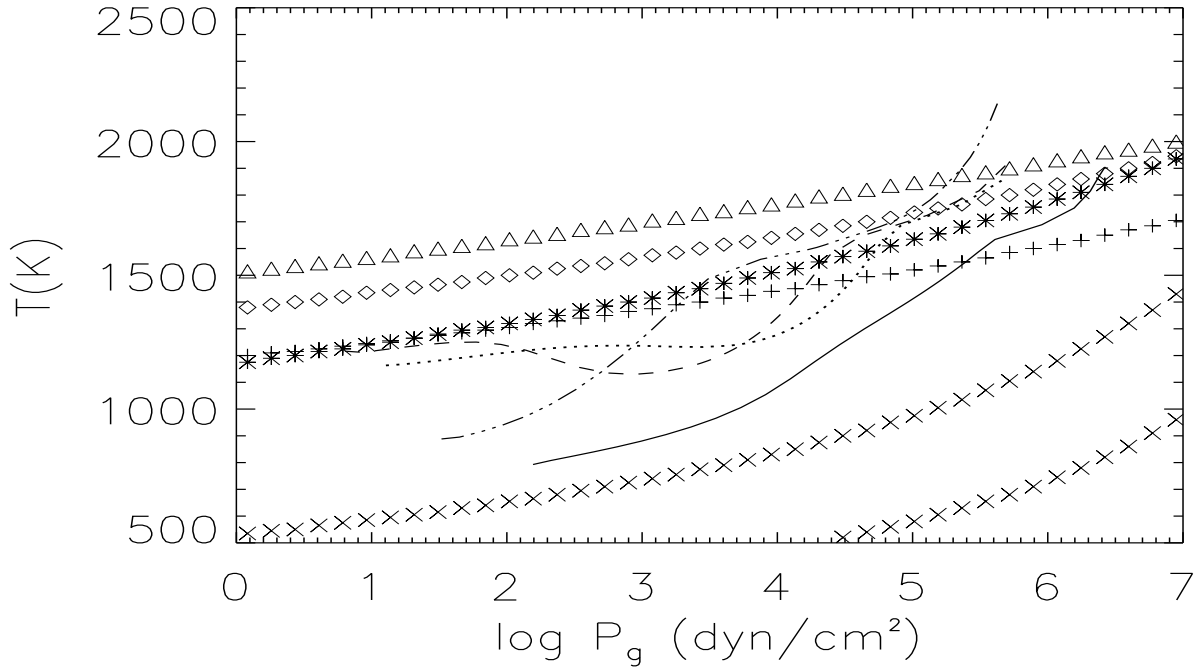}{3.5in}{0}{100}{80}{-85}{-300}}
\caption{Temperature-pressure profiles for four different 51 Peg b
models. The models shown correspond to a homogeneous
MgSiO$_3$-Al$_2$O$_3$-Fe cloud with particles of $\overline{r}=0.01~\mu$m
(dotted), 0.1~$\mu$m (dashed),
and 10~$\mu$m (dot-dashed line). The solid line is a model with
only MgSiO$_3$ clouds with particles of $\overline{r}= 0.01~\mu$m.
The symbols show the condensation curves of CaTiO$_3$
(triangles), Al$_2$O$_3$ (diamonds), Fe (*), MgSiO$_3$ (+), the
CO/CH$_4$ equilibrium curve (upper {$\times$}), and the
N$_2$/NH$_3$ equilibirum curve (lower {$\times$}).
The metallicity is [Fe/H] = +0.21, corresponding to that
of 51 Peg~A.}
\label{fig:TempProfile}
\end{figure}

The temperature-pressure profiles (which
are the basis for the emergent spectra) also vary depending
on the type and size distribution of condensates in 
the planet atmosphere. 
Figure~\ref{fig:TempProfile} shows the temperature-pressure
profiles of the four models shown in Figure~\ref{fig:spectra},
together with the equilibrium condensation curves.
In contrast to the T dwarfs which do not need
clouds to be modeled, the irradiated CEGPs have heated
upper atmospheres that bring the temperature closer
to the equilibrium condensation curves so are more likely to have clouds
near the top of the atmosphere.
In Figure~\ref{fig:TempProfile}
the model with particle mix with $\overline{r}=0.1~\mu$m
(dashed line) is highly absorbing and 
results in a temperature inversion in the upper atmosphere layers.
The model with cloud with particles with $\overline{r}=10~\mu$m
is much less absorbing and
results in a cooler temperature in the upper atmosphere layers.
A highly reflective model (solid line)  shows that much less heating
occurs when molecules such as H$_2$O are the primary absorbers.
With the clouds at low
pressure, at $10^3$ -- $10^4$ dyne~cm$^{-2}$, the equilibrium
condensation curves for MgSiO$_3$ and Fe are close together
so a cloud mix of both particles could exist.
Even if the uppermost cloud dominates the reflected light curve
and spectra, the heating from lower cloud layers such
as Al$_2$O$_3$ are important and do alter the temperature-pressure
profile. The $T_{\rm eff}$s of these models range from 1170~K to 1270~K.

Another interesting consequence of the irradiative heating,
evident from Figure~\ref{fig:TempProfile},
is the proximity of the temperature-pressure profiles to the
CO/CH$_4$ equilibrium
curve. As noted in Goukenleuque et al. (1999) CO is expected
to dominate over CH$_4$, but CH$_4$ is abundant enough
to produce absorption bands. We have found that the strength
of the CH$_4$ features is sensitive to the upper atmosphere
temperature which is in turn dependent on the amount
of irradiative heating. Thus the CH$_4$ bands are a good
temperature diagnostic. (The CH$_4$ bands are also useful, but less sensitive,
as a pressure diagnostic (Seager 1999)).

Alternate approaches to modeling the temperature-pressure
profiles have been taken. Marley et al. (1999)
consider the temperature-pressure profile
of an isolated object of the same effective temperature.
Sudarsky et al. (1999) use ad hoc modified isolated temperature-pressure
profiles (based on the temperature-pressure profiles in Seager \& Sasselov 1998),
to simulate heating, instead of computing irradiative
heating. As a result, the temperature gradient 
is much steeper because most of the heat comes
from the bottom of the atmosphere.
These models have clouds at the 1 bar level,
near the bottom of the atmosphere.
One consequence of this assumption
is the strength of the K~I and Na~I absorption.
Because of the clear atmosphere down to the 1 bar level,
the K~I and Na~I resonance lines are extremely
pressure broadened and absorb essentially all incoming
optical radiation redward of 500~nm.
This is in contrast to the spectra shown in
Figure~\ref{fig:spectra} where the K~I and Na~I resonance lines are relatively
narrow --- 
the deep pressure zones where the broad line wings are formed are
not sampled.

Although our approach is to compute the temperature-pressure
profiles and reflected light in a consistent manner,
we emphasize that in general there are many uncertainties
in current CEGP models
including photochemistry, cloud assumptions, and heat redistribution by winds.
More specific to our models is the internal heat assumptions.
For numerical reasons we must assume
a lower boundary condition to our atmosphere in the form
of a net flux coming from the planet interior. The assumption we have made
is a net flux of approximately 1/10th of the absorbed flux.
This may be too high (T. Guillot, private communication),
and using a much lower value would produce a more
isothermal atmosphere at the highest pressures in
our models. More work is needed
to understand the 3D heating redistribution in CEGPs.
Importantly, the lower boundary
condition has little effect on the upper atmospheric temperature and 
the reflected light curves.

\subsection{Condensates and the Scattering Asymmetry Parameter}
\label{sec-g}
\begin{figure}
{\plotfiddle{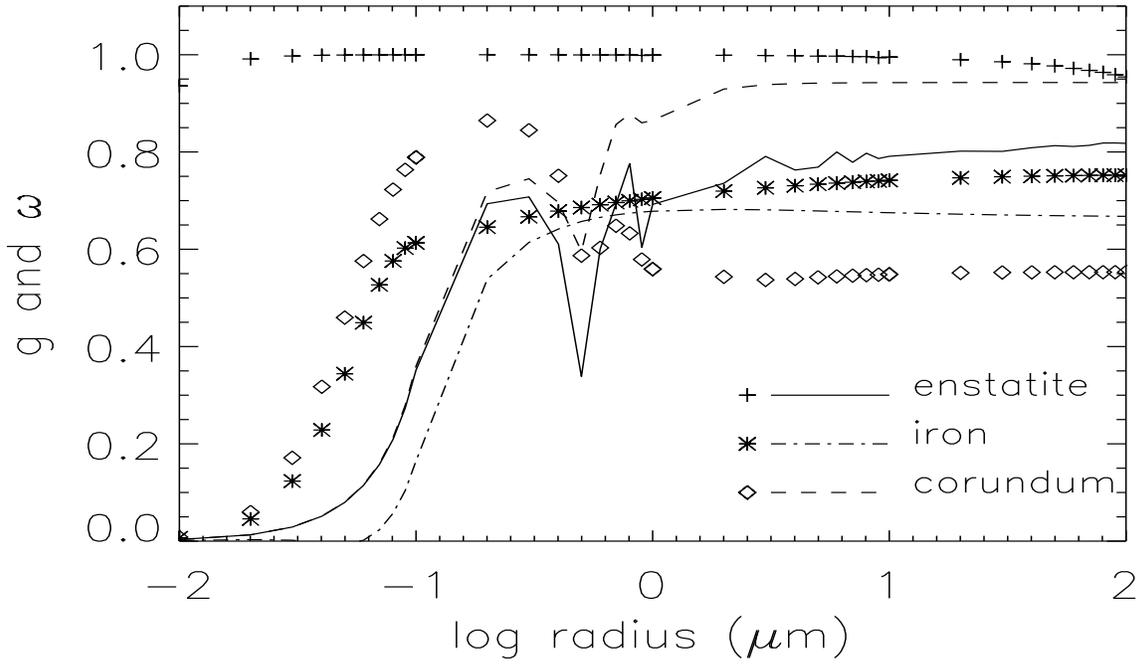}{2in}{0}{100}{80}{-85}{-300}}
\caption{Scattering asymmetry parameter (lines) and single scattering
albedo (symbols) for the three condensates used in this study. See discussion in
text for details.}
\label{fig:g}
\end{figure}

The shapes of the CEGP reflected light curves depend on the
absorptivity and directional scattering probability
of the condensates.
Figure~\ref{fig:g} shows the scattering asymmetry parameter and the
single scattering albedo at 5500 \AA~ for the three condensates
considered as a function of particle size. The scattering asymmetry
parameter $g$ is defined by
\begin{equation}
g = <\cos {\theta_S}> = \int_{4 \pi} \cos {\theta_S} P^{11}
\frac{d\Omega}{4 \pi},
\end{equation}
where $\theta_S$ is the scattering angle and $P^{11}$ is the phase
function (see e.g. Van de Hulst 1957). The scattering phase function
is the directional scattering probability of condensates
(see Figures~\ref{fig:phasefn1} and~\ref{fig:phasefn0.1c}), and should
not be confused with the planetary phase function introduced in \S\ref{sec-analytical}
The scattering asymmetry parameter
$g$ varies from -1 to 1 and is 0 for isotropic scattering. The higher
$g$ is, the more forward throwing the particle.
The curves for $g$ and $\tilde{\omega}$
in Figure~\ref{fig:g} can predict, or help interpret, the light curves.
Small particles compared to wavelength scatter as Rayleigh
scattering; $g=0$ in this case, where the forward and backward
scattering average out. This is seen for particles with $r=0.01$~$\mu$m
(along the $y$ axis). In addition, Al$_2$O$_3$ and Fe have
$\tilde{w}=0$ for $r=0.01$~$\mu$m,
so for these small particles absorption dominates over all scattering
and the light curves will show little variation. For particles with
$r=0.1$~$\mu$m more scattering will occur than for $r=0.01$~$\mu$m;
$g=0.2$ and $\tilde{w}$ is high. In this case, scattering is not too
forward
throwing, and the probability of scattering over absorption is
high. For particles with $r=10~\mu$m, both $g$ and $\tilde{w}$ are high.
High $g$
means the particle will scatter light preferentially in the forward
direction. Coupled with high $\tilde{w}$, the photons will multiply
scatter
forward into the planet resulting in little reflected light. These
effects will be partially borne out in the light curves shown in the next
section.

The single scattering albedos and the asymmetry parameter curves are
generally similar for large $r$ for the three particles considered
because the
parameters --- indeed light scattering in general --- are determined
largely by the particle size compared to the wavelength of light.
The curves are different from each other because of the different
nature of the particles,
specifically the real and complex indicies of refraction. At 5500~\AA,
Fe has a very high complex index of refraction, while that of
MgSiO$_3$ is essentially zero.
The variations for a given curve, notably at $r=0.5$~$\mu$m, are
interference effects between diffracted light rays and rays that
refract twice through the particle. This effect gets damped out for
absorptive particles (e.g. Fe). For a concise discussion of asymmetry
parameters and phase functions see Hansen \& Travis (1974).

Although we have chosen the dominant solids expected at
the relevant temperatures and pressures from equilibrium 
calculations,
it is certainly possible that nature has provided CEGPs with a
different condensate size distribution,
and different condensate particles and shapes than the ones used
here. This will be investigated in future work.

\subsection{Visual Photometric Light Curves and Polarization}
\label{sec-visual}
In the next few subsections we present the photometric light curves
and fractional polarization curves.
The results from our Monte Carlo scattering code give a planet-star
flux ratio, and we use equation~(\ref{eq:dm}) to convert to variation in
$\Delta m$, but where $\epsilon \phi(\Theta)$ is the planet-star
flux ratio ($\Theta$ is defined below).
The results from the Monte Carlo scattering code also
give the percent polarization, $Pol$, and we convert this to fractional
polarization of the system as described in \S\ref{sec-analytical},
by $P_{\rm frac} = \epsilon \phi(\Theta) Pol$, with the flux ratio
in place of $\epsilon \phi(\Theta)$.

Figures~\ref{fig:lc0.01}a, ~\ref{fig:lc0.1}a, ~\ref{fig:lc1}a, and
~\ref{fig:lc10}a show the 5500~\AA~light curves with orbital
angle $\Theta$
for the four cases of particles with mean radius $\overline{r}=$ 0.01, 0.1, 1,
and 10~$\mu$m. Here we use orbital angle instead of orbital phase used
for radial velocity measurements because an angular variable is more
convenient for the analysis. We define $\Theta$ as
the angle in the orbital plane of the planet and star.
With this definition, an orbital angle of
0$^{\circ}$ occurs when the planet is farthest from Earth, and an
orbital angle of 180$^{\circ}$ occurs when the planet is between Earth
and the star.
In addition, for $i=90^{\circ}$ orbital angle and phase angle are
equivalent, and orbital angle 0$^{\circ}$ corresponds to orbital phase
of 3/4 (used in radial velocity measurements).
In each figure the first curve is for inclination
$i=90^{\circ}$, which for all CEGPs except HD~209458~b,
has already been excluded by transit non-detections. The other curves
are for $i=82^{\circ}$,
$i=66^{\circ}$, $i=48^{\circ}$, and $i=21^{\circ}$, none of which can be
excluded by transit non-detections. The $y$ axis scale
differs among
the different figures. Transits occur only for $i \geq
\theta_T$
($\theta_T = 83.3^{\circ}$ for 51 Peg b
with $R_*=1.16$~$R_{\odot}$, $R_P=1.2$~$R_J$, and $D=0.051$~AU), and
within $\Theta = 180^{\circ} \pm (90-\theta_T)$. They
are barely visible on these
figures, since transits darken rather than brighten the
light of the system. In addition, the transit light curves are on
the order of millimagnitudes, $\sim$2 orders of magnitude greater
than the reflected light effect. Nevertheless the start of the drop in
the light curve for $i=90^{\circ}$
is shown at $\Theta = 173.7^{\circ}$ and $\Theta = 186.3^{\circ}$
at first and fourth contact
respectively. 
Similarly,
for $i=90^{\circ}$ --- and only for $i> \theta_T$ --- the
reflected
planetary light is not visible as the planet
goes behind the star at $\Theta = 360^{\circ} - (90^{\circ}-\theta_T)$, and reemerges
at $\Theta = (90^{\circ}-\theta_T)$; that area is
shaded in on the figures.

Figures ~\ref{fig:lc0.01}b, ~\ref{fig:lc0.1}b, ~\ref{fig:lc1}b, and
~\ref{fig:lc10}b show the fractional polarization with orbital angle for
the four
cases of mean particle radius 0.01, 0.1, 1, and 10 $\mu$m. Inactive
solar-type stars are very weakly polarized, on
the order of a few $\times$ $10^{-2}$ percent, so we treat the incoming
light as unpolarized.  We have plotted the fractional linear
polarization of the system as described in \S\ref{sec-analytical}, so
that the polarization is modulated by the amount of scattered light,
which peaks at an orbital angle of zero. Circular polarization is a
secondary effect, and smaller than the errors in our scattering
simulation.

\subsubsection{Particles with $\overline{r}=0.01~\mu$m}
\label{sec-size0.01}
\begin{figure}
{\plotfiddle{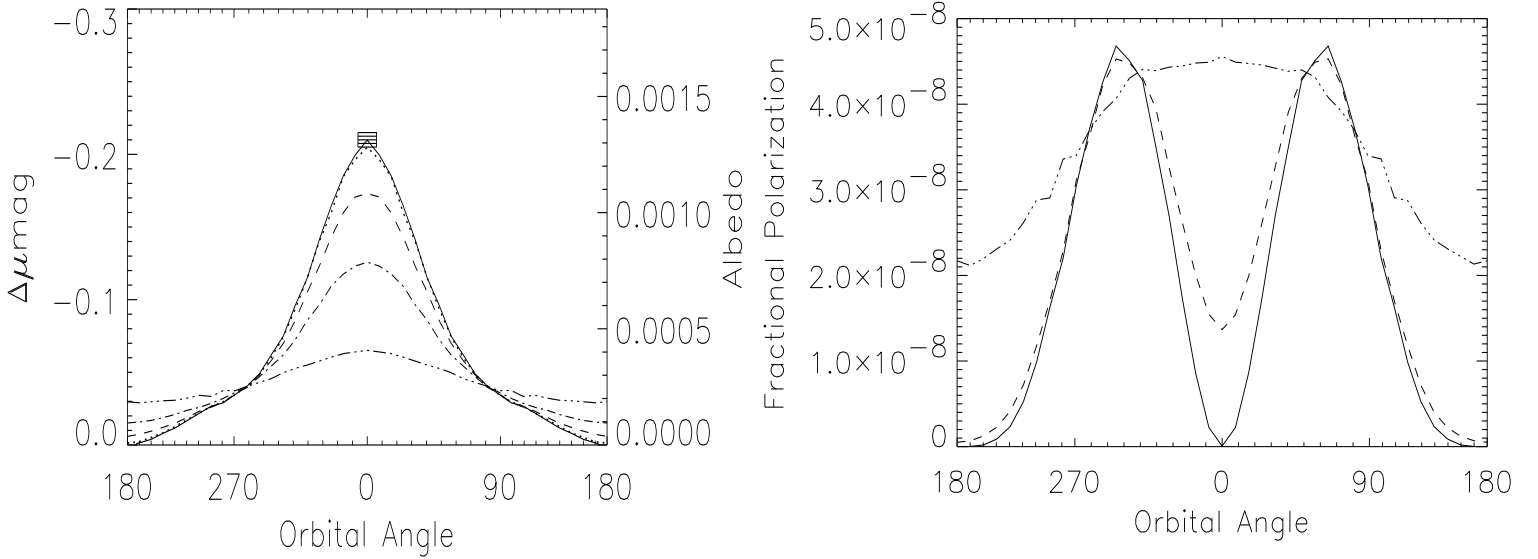}{2in}{-90}{80}{80}{-75}{310}}
\caption{Light curves and fractional polarization for particles with
$\overline{r}=0.01~\mu$m. The lines correspond to different inclinations: solid =
90$^{\circ}$, dotted = 82$^{\circ}$, dashed = 66$^{\circ}$, dash-dot =
48$^{\circ}$, dash-dot-dot-dot = 21$^{\circ}$. The
hatched area is not observable; it represents the orbital angles at which
the star is directly in front of the planet, which only occurs for $i>
\theta_T$. For clarity the hatched area is not
shown for the polarization fraction, and only
three of the inclinations are shown. Fractional polarization at
$i=21^{\circ}$ is noisier than at other inclinations because
few photons scatter into these phase angles, as indicated
by the light curve (left panel). }
\label{fig:lc0.01}
\end{figure}

\begin{figure}
{\plotfiddle{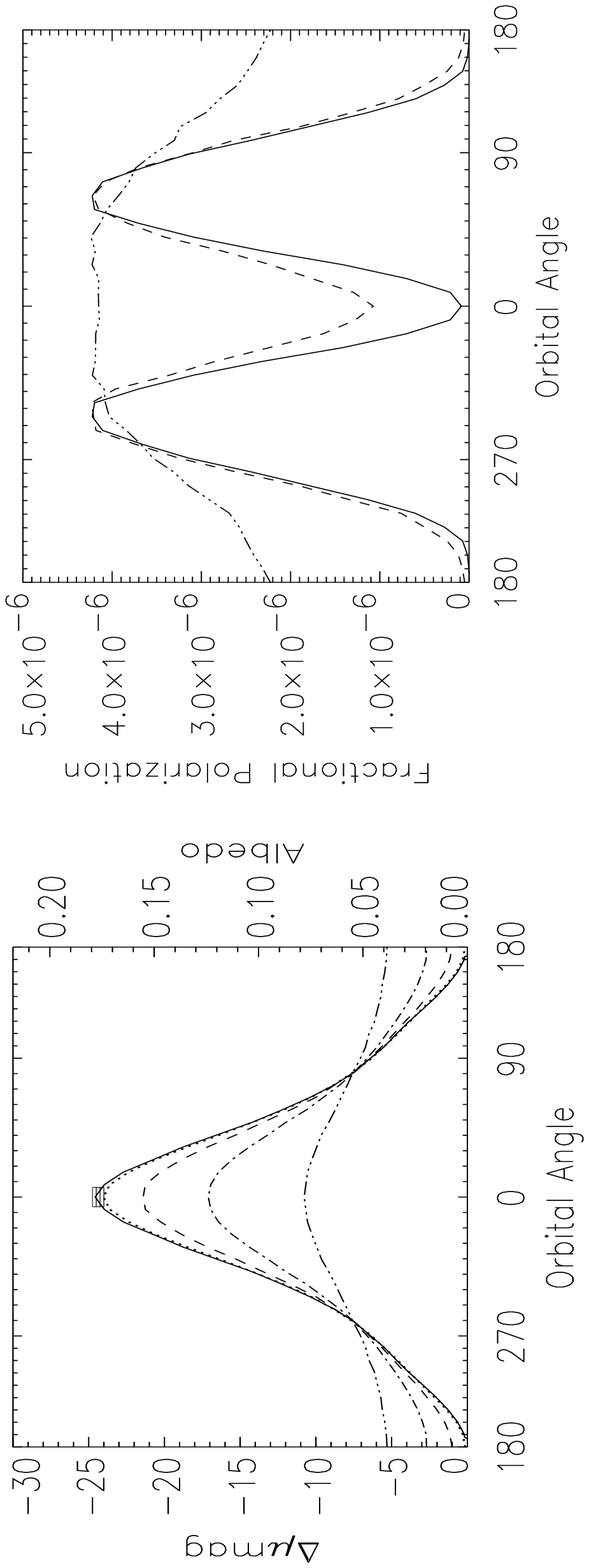}{2in}{-90}{80}{80}{-75}{310}}
\caption{Light curves and fractional polarization for particles with $\overline{r}=
0.1~\mu$m. The curves are for the same inclinations as in Figure
\ref{fig:lc0.01}.}
\label{fig:lc0.1}
\end{figure}

\begin{figure}
{\plotfiddle{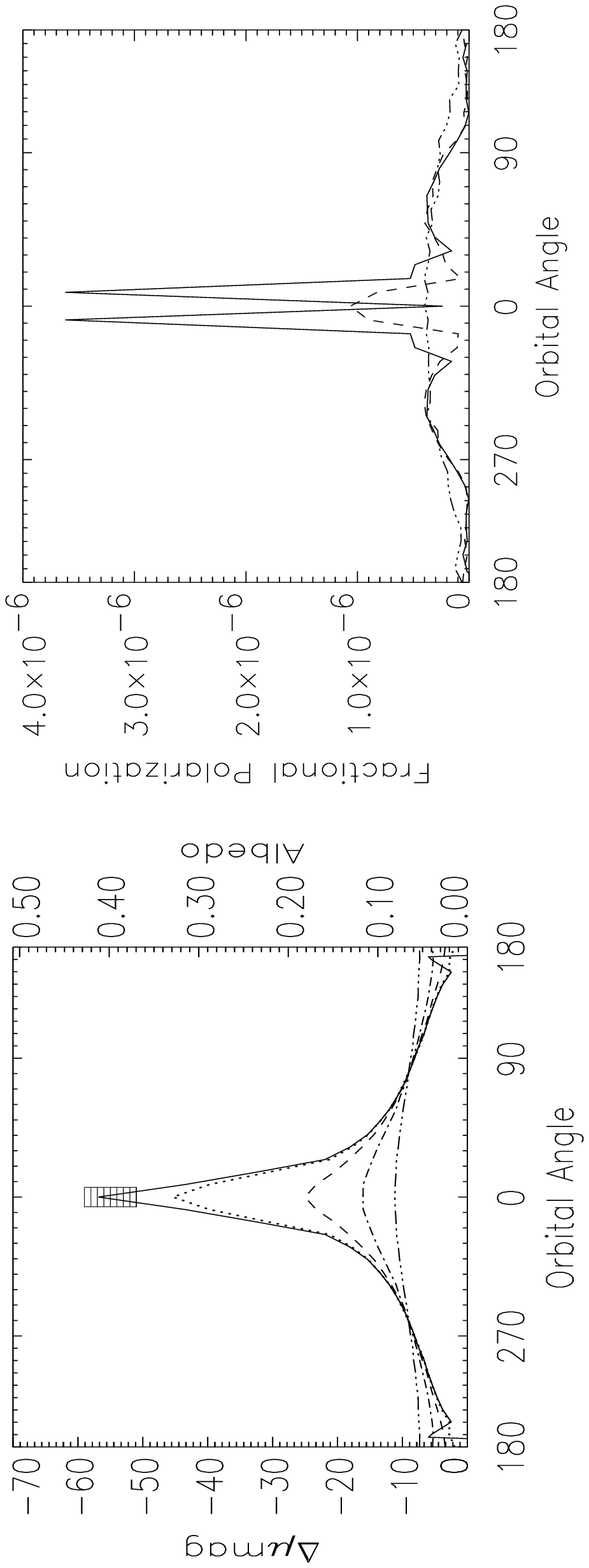}{2in}{-90}{80}{80}{-75}{310}}
\caption{Light curves and fractional polarization for particles with $\overline{r}= 
1~\mu$m. The curves are for the same inclinations as in Figure
\ref{fig:lc0.01}.}
\label{fig:lc1}
\end{figure}

Figure~\ref{fig:lc0.01}a shows the light curve for particles with $\overline{r}=0.01~\mu$m,
the case of very small particle size compared to
wavelength. The amount of scattered light is tiny due to the high
absorptivity of Fe and Al$_2$O$_3$, as described in \S\ref{sec-g} That
small particles obey Rayleigh scattering can be seen from the smooth,
Rayleigh-like shape of the light curve. 
Rayleigh scattering produces more
backscattering and results in a slightly different light curve from
isotropic scattering.
The Rayleigh scattering phase function $P_{Ray} =
3/4(1+\cos^2 \theta_S)$ (where $\theta_S$ is the scattering angle), whereas
for purely isotropic scattering
$P_{iso} = 1$. At $\Theta=0^{\circ}$ and $i=90^{\circ}$, $P_{Ray}$ = 1.5, compared to
$P_{iso} = 1$, and at $\Theta=90^{\circ}$ and $i=90^{\circ}$, $P_{Ray}$ = 0.75
which is smaller than $P_{iso}=1$. 

Rayleigh scattered light is maximally polarized at a scattering angle
of 90$^{\circ}$. Figure~\ref{fig:lc0.01}b shows the fractional
polarization of the CEGP system (described in \S\ref{sec-analytical}),
i.e. the ratio of polarized light to total white light of the star + planet.
Because absorption is so
high for this case, photons that exit the sphere have singly
scattered, and the scattered light is 100\% polarized. Even so, 
the fractional polarization is tiny because the amount of scattered
light is very small compared to the unpolarized stellar flux.
Figure~\ref{fig:lc0.01}b also shows that different inclinations have
the same polarization peaks, but with smaller amplitudes.
Fractional polarization at $i=21^{\circ}$ is noisier than
at other inclinations, since less radiation scatters
into these phase angles, as indicated by the light curve
in Figure~\ref{fig:lc0.01}a.

We also ran simulations with MgSiO$_3$ particles
as the only condensate present in the planetary atmosphere (not shown in
the figures), to investigate the lightcurves without the highly
absorbing Fe and Al$_2$O$_3$ condensates (see Table~2).
For $\overline{r}=0.01$~$\mu$m
at $i=90^{\circ}$,
the light curve peaks at 25 $\mu$mag, which corresponds
to a geometric albedo $p=0.18$. Although $\tilde{\omega}$
is relatively high for this case, multiple scattering
makes the resulting geometric albedo much smaller than
for single scattering because it gives more chance for absorption.
For the case where only MgSiO$_3$ with particles of $\overline{r}=0.01$~$\mu$m is
present, the polarization fraction peaks at
$5.5 \times 10^{-6}$, which is almost 2 orders of magnitude higher than the
MgSiO$_3$-Fe-Al$_2$O$_3$ mix. However, it is still
below current detectability limits.

\subsubsection{Particles with $\overline{r}=0.1~\mu$m}
\label{sec-size0.1}

The light curves for particles with $\overline{r}=0.1~\mu$m, shown in
Figure~\ref{fig:lc0.1}a, are similar to those for $\overline{r}=0.01~\mu$m, but
have a much larger amplitude.
This can also be seen from Figure~\ref{fig:g} which shows that for visual
wavelengths the single scattering albedos are higher than those for
$\overline{r}=0.01~\mu$m, and the scattering
asymmetry parameter is only 0.2, which is reasonably isotropic.

Polarization, shown in Figure~\ref{fig:lc0.1}b, is also similar
to the $\overline{r}=0.01~\mu$m case, with a much larger amplitude.
Because the particles are still somewhat small
compared to the wavelength of light, the scattering is largely
Rayleigh scattering and the
peaks are similar to those in Figure~\ref{fig:lc0.01}b. However, not
shown is that the scattered light is 55\% polarized.

For this case of particles with $\overline{r}=0.1~\mu$m, $\tilde{\omega}$ of Fe
dominates. As seen from 
Figure~\ref{fig:g}, if $\tilde{\omega}$ from MgSiO$_3$ or Al$_2$O$_3$
dominates instead, the light
curve will have a higher amplitude. For example, for a model with pure MgSiO$_3$ clouds,
the $i=90^{\circ}$ light curve peaks at 95 $\mu$mag, which corresponds
to a geometric albedo $p=0.69$, and the fractional polarization
peaks at $8.6 \times 10^{-5}$.
The $i=21^{\circ}$ light curve peaks at 42 $\mu$mag.
This case has the highest reflectivity of all of our models.

\subsubsection{Particles with $\overline{r}=1~\mu$m}
\label{sec-size1}
\begin{figure}
{\plotfiddle{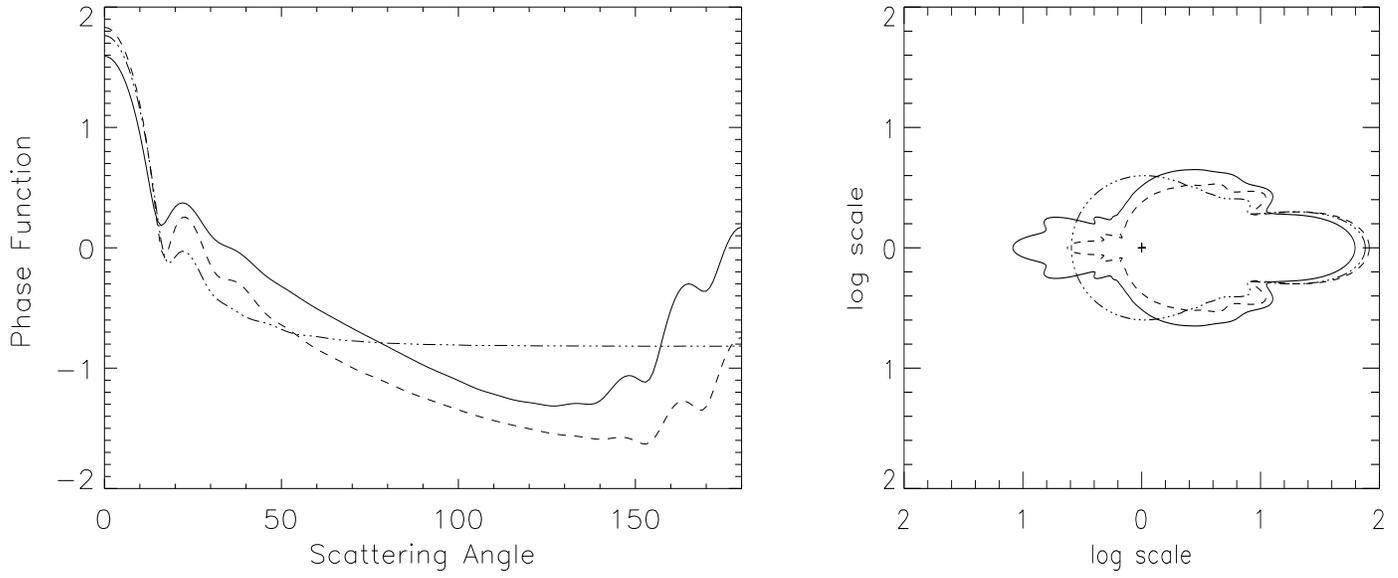}{1.9in}{-90}{80}{80}{-75}{340}}
\caption{The phase functions and polar diagrams for the three condensates
with $\overline{r}=1~\mu$m. The solid line is MgSiO$_3$, the dash-dot line Fe,
and the dashed line Al$_2$O$_3$. In the polar diagram the light is
incoming from the left and the condensate particle is marked by the cross.
The axes on the polar diagram are in a log scale, the units are
dimensionless and only the relative numbers are important.}
\label{fig:phasefn1}
\end{figure}

For particles with $\overline{r}=1~\mu$m, which are larger than visual
wavelengths, the light curve shown in Figure~\ref{fig:lc1}a shows
effects both from forward throwing and from a narrowly peaked
backscattering function.
Figure~\ref{fig:phasefn1}a shows the phase function for
the three different condensates, plotted with scattering angle $\theta_S$.
Although the phase function represents single
scattering, it can be used to interpret the light curve which
arises from multiple scattering.
The narrow backscattering (at $\theta_S=180^{\circ}$) is responsible for the
narrow peak in the light curve: there is a high probability of
backscattering but only for a narrow angular range.
The high probability for forward throwing means
that photons are likely to be forward scattered into the atmosphere where they
will be absorbed; this is the cause for the otherwise reduced light
curve (in the ``wings'') compared to the Rayleigh-shaped light curves in
Figures~\ref{fig:lc0.01}a and ~\ref{fig:lc0.1}a.

\begin{figure}
{\plotfiddle{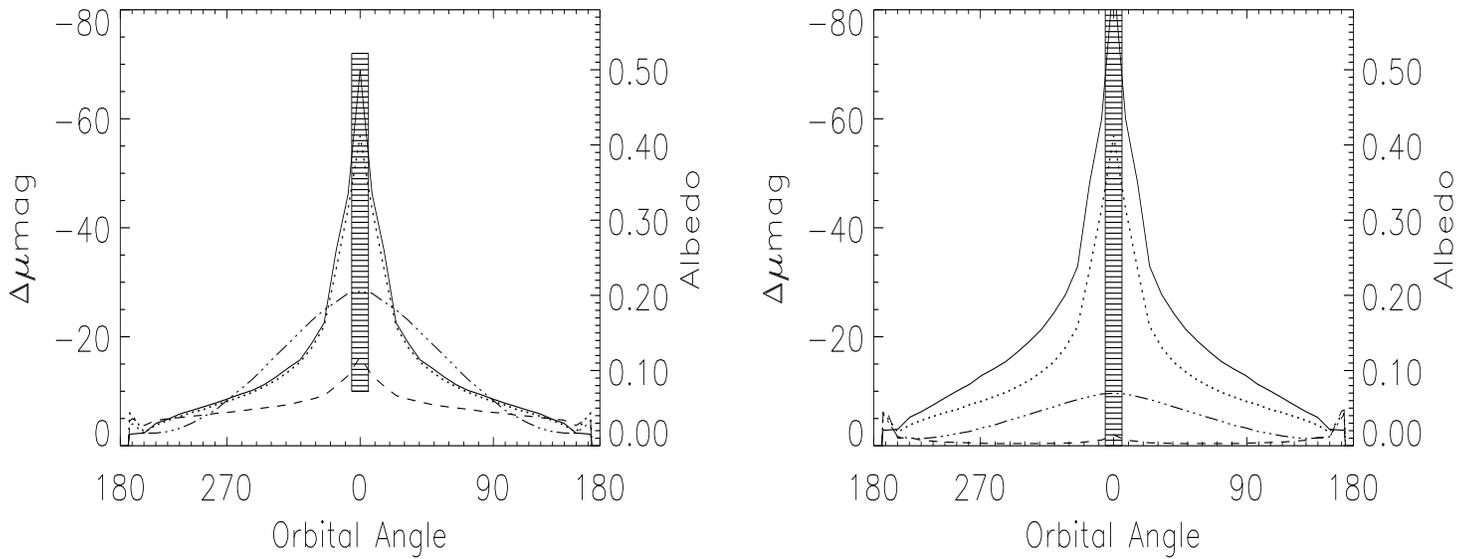}{2in}{-90}{80}{80}{-75}{310}}
\caption{Light curves for $i=90^{\circ}$ for individual condensate
particles with $\overline{r}=1~\mu$m. The dotted lines are the light curves from
the MgSiO$_3$-Fe-Al$_2$O$_3$ mix, the solid line is MgSiO$_3$, dot-dash
is Fe, and dashed is Al$_2$O$_3$. Figure~\ref{fig:lc1inddust}a shows
the light curves as if each condensate was the only one present in the
atmosphere. Figure~\ref{fig:lc1inddust}b shows the light curves for
the phase function of each condensate, but with the same total opacity
as in the MgSiO$_3$-Fe-Al$_2$O$_3$ mix.}
\label{fig:lc1inddust}
\end{figure}

\begin{figure}
{\plotfiddle{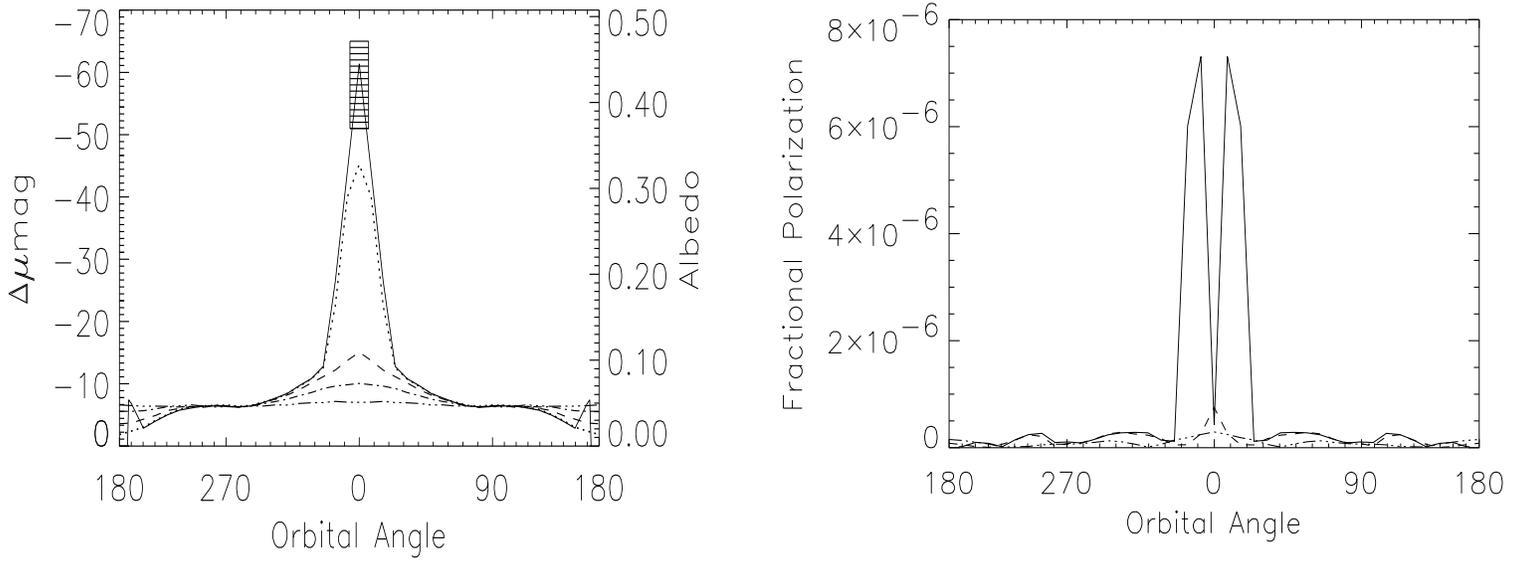}{2in}{-90}{80}{80}{-75}{310}}
\caption{Light curves and fractional polarization for particles with 
$\overline{r}=10~\mu$m. The curves are for the same inclinations as in Figure
\ref{fig:lc0.01}. The rise in the light curves at $\Theta >
160^{\circ}$ is from light forward scattering through the upper
atmosphere. Part of the transit light curve is visible near
$\Theta=180^{\circ}$.}
\label{fig:lc10}
\end{figure}

We also plot the phase function as a polar diagram
in Figure~\ref{fig:phasefn1}b, where the light is incoming
from the left, and the condensate particle is marked at the origin.
Figures~\ref{fig:phasefn1}a and \ref{fig:phasefn1}b also show that
the three condensates have different amounts of backscattering, indeed
slightly different phase functions overall. In
Figure~\ref{fig:lc1inddust}a we plot the three light curves from
each of the condensates, considering that each condensate is the only one
present in the atmosphere. This shows that the $\tilde{\omega}$ of
MgSiO$_3$ dominates, as compared to particles with $\overline{r}=0.01~\mu$m where Fe
dominates, and also that each condensate has unique properties.
Both Fe and Al$_2$O$_3$ have very forward
throwing phase functions without a strong backward peak; the
result is that incoming light is forward scattered into the atmosphere where it
is not likely to contribute to reflected light. As a result,
their light curve amplitudes are much smaller than MgSiO$_3$'s; Al$_2$O$_3$ alone
would not even be detectable by the planned microsatellites.
The same comparison for particles with $\overline{r}=0.01~\mu$m
and $\overline{r}=0.1~\mu$m, where the phase functions and light curves are more isotropic,
reveals that the main effect of each condensate is mostly a change in
magnitude. This is because the particles are smaller than the wavelength of light,
and to first order light is Rayleigh scattered.

The phase function of MgSiO$_3$, the solid curve in
Figure~\ref{fig:phasefn1}a, is typical of those for
spheres of size
parameter $x \ge $ 1, with complex index of refraction $n_i \sim 0$ (see
Hansen \& Travis 1974), where $x=2 \pi r/\lambda$.
The forward throwing is caused by
diffraction of light rays around the particle and depends on the
geometrical cross-section of the particle, and so would also occur for
non-spherical particles.
The backward peak, known as the
``glory'', is specific to spherical particles and
is related to interfering surface waves on the particle sphere.
For absorbing particles
(with high $n_i$, such as Fe), this effect
is damped out, as seen from the Fe
phase function in Figure~\ref{fig:phasefn1}a, which shows no rise
towards $\theta_S=180^{\circ}$. For randomly oriented
axi-symmetric spheroids, the backward peak would not be
as severe (Mishchenko et al. 1997). The phase function of randomly
oriented
axi-symmetric spheroids depends on
the distribution of both particle axis sizes and particle
orientation. To estimate the light curve from a reduced backscattering
peak, we assume all of the condensates scatter
like Fe, and in another case all like Al$_2$O$_3$. In
other words, we use the same total opacity of the
MgSiO$_3$-Fe-Al$_2$O$_3$ mix. The
resulting
light curves are shown in Figure~\ref{fig:lc1inddust}b.
Although their peaks are much lower than the strong backscattering
case, with the exception of Al$_2$O$_3$ this
variation is still detectable by the upcoming microsatellite
missions.

Because the CEGPs are very close to their parent stars, light rays
hitting the planet may not be well approximated as plane-parallel.
We use the correct angular distribution of ${\rm tan}^{-1}(R_*/D)$
(6$^{\circ}$ for 51 Peg b).
The main effect from using this angular
distribution compared to plane-parallel rays is that the
backscattering peak is reduced by 12\%, because the
backscattering phase function peaks sharply at $180^{\circ}$.
A more minor improvement is that the lower inclination light curves are a
few percent lower.
For isotropic or Rayleigh scattering we find no difference in the
light curve from using either plane-parallel rays or the correct
angular distribution of incoming radiation.
Because ${\rm tan}^{-1}(R_*/D)$ is such a small angle,
isotropic irradiation, used in atmosphere codes that treat feedback
from the star's own corona, is not accurate.

The polarized light curve shown in Figure~\ref{fig:lc1}b is very
different from the Rayleigh scattering polarization curves for
$\overline{r}=0.01~\mu$m and $\overline{r}=0.1~\mu$m.
The polarization is more complex than Rayleigh scattering as the
light rays reflect from and refract through the particles, and the
scattered light rays interfere.
In addition, the polarization from each
condensate is different, since polarization depends in part on
the index of refraction,
which is very different for each of the three condensates in this
study. The peak of the polarization has a similar
peak to the light curve, since fractional polarization is plotted
which follows the scattered light. Polarization of the scattered light
alone shows a smaller central peak and additional smaller peaks at
10$^{\circ}$ and at 70$^{\circ}$ for
$i=90^{\circ}$.

\subsubsection{Particles with $\overline{r}=10~\mu$m}
\label{sec-size10}
Light curves from particles with $\overline{r}=10~\mu$m, shown in
Figure~\ref{fig:lc10}a, are similar to the light curves from $\overline{r}=1~\mu$m
particles, but with a more pronounced effect from strong forward throwing
and an even more narrowly peaked backscattering probability.
Outside of the backscattering peak,
the light curve is much smaller due
to the forward throwing effects discussed above.
For $i=90^{\circ}$ and $\Theta=160^{\circ}$, there is a rise in the
light curve just before the transit. This is from light that enters
the atmosphere near the limb, and scatters through the top of the
atmosphere due to the high forward throwing nature of the large
particles. For particles with $\overline{r}=10~\mu$m, the polarization is very similar to
the $\overline{r}=1~\mu$m
case, but with a greater peak from the greater amount of
backscattering, and an otherwise lower amplitude from the higher
forward throwing.

\subsection{U, B, V, R Photometric Light Curves and Polarization}
In this section we compare the light curves and polarization at the U, B,
V, and R effective wavelengths (U = 3650 \AA, B = 4400 \AA, V =
5500~\AA, R = 7000~\AA).
The incoming stellar flux has many spectral features over the
wavelength range of a band, but the same features are reflected by the
planet (Figure~\ref{fig:spectra}); the light curve depends on the flux
ratio and the features cancel out.
However, for a second
order calculation absorption by alkali
metal line wings (in the planetary atmosphere) or 
the opacity variation from condensates within a color band may play a
role.

In general the light curves are a function of the opacity and of the
phase function. The opacity effects include density effects and the
single scattering albedo. Photons travel into
the atmosphere and encounter condensates (or atoms or molecules) which
will scatter or absorb them. If the photons
scatter, they will scatter according to the condensate phase
function. A phase function that preferentially scatters photons into
the forward direction will generate a very different light curve than
a Rayleigh phase function, as shown in \S\ref{sec-visual}
The phase function for given optical constants depends on the
size parameter $x = 2 \pi r / \lambda$, and the condensate index of
refraction.

\begin{figure}
{\plotfiddle{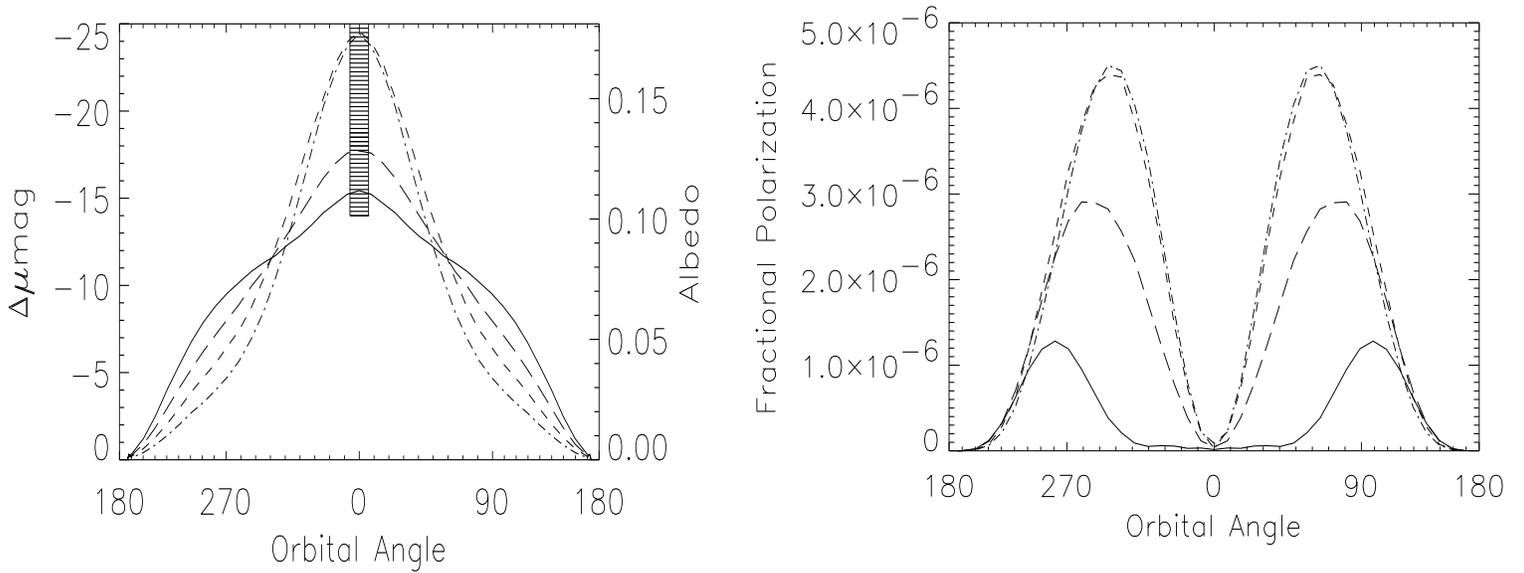}{2in}{-90}{80}{80}{-75}{310}}
\caption{Light curves and fractional polarization at $i=90^{\circ}$
for condensates with 
$\overline{r}=0.1~\mu$m, for U (solid), B (long dash), V (dash), and R(dot-dash).}
\label{fig:lcc0.1}
\end{figure}

\begin{figure}
{\plotfiddle{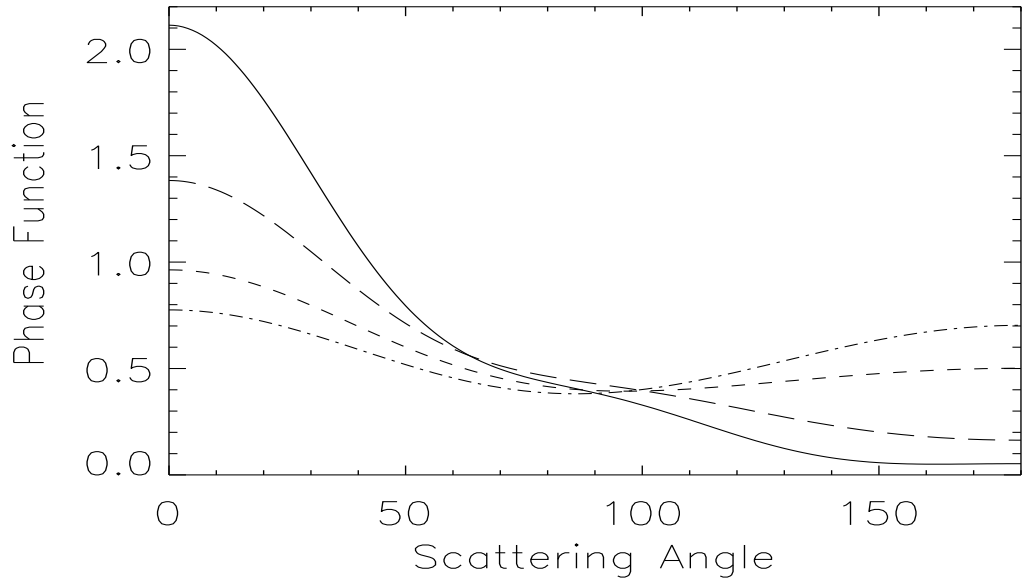}{2.5in}{0}{90}{70}{-85}{-255}}
\caption{Phase function for Fe, for particles with
$\overline{r}=0.1~\mu$m, for U (solid), B (long dash), V (dash), R(dot-dash).}
\label{fig:phasefn0.1c}
\end{figure}

Figure~\ref{fig:lcc0.1}a shows the light curves
of the 51 Peg b system for particles with $\overline{r}=0.1~\mu$m. The main
difference between the colors is caused by the different size
parameters: different wavelengths
of light for a fixed particle size. For example, $x=0.90$ at R but
$x=1.72$ at U. 
The effects of this are seen in the
light curves: R has a Rayleigh-scattering-like light curve compared to U.
The phase
functions for the 4 colors for Fe are shown in Figure~\ref{fig:phasefn0.1c}.
As mentioned in \S\ref{sec-size0.1}, Fe
opacity dominates this case of particles with $\overline{r}=0.1~\mu$m. In fact, the U
light curve is close to the shape of the Fe-only light curve for
$\overline{r}=1~\mu$m particles at V (dashed line in
Figure~(\ref{fig:lc1inddust}a)). 
Because of the $x$ dependency of the phase functions, the different
colors for a fixed size go through the same shapes as shown for the V
light curves which describe fixed wavelength for a varying particle size.
However, there are differences due to opacity variation
with color.

Figure~\ref{fig:lcc0.1}b shows the polarization fraction for
$\overline{r}=0.1~\mu$m. Effects from both phase function and
opacity contribute. 
The polarization peak of each color is at a different angle, due to
the different indicies of refraction of the particles at different colors.
The difference in the polarization peaks
is much greater than the difference in the light
curve peaks between the colors. The polarization curves reflect
the higher asymmetry in the scattering in U and the greater absorption at
this wavelength.

\subsection{Cloud Layers}
\label{sec-CloudLayers}
The light curves and polarization curves presented in this paper
have been computed under the assumption that the clouds are
not in layers, but that above a given
equilibrium condensation curve
all of the gas condenses into solids and the particles are suspended uniformly
in the atmosphere. 

There are two consequences of finite cloud layers.
The first difference is that stratification will separate different
condensates into different layers, where the light curve
and flux signature will come largely from the cloud closest
to the top of the atmosphere.
In contrast to our models of vertically homogenous clouds,
a cloud confined to one pressure scale height with the same solid
mass fraction would have a higher optical depth.
With a high optical depth  the reflected light
cloud signature is even more likely to resemble the uppermost
cloud only than the situation of a homogenous mix.

The second difference is that the gaseous scatterers and absorbers
above the cloud layer could play a role. The dominant gaseous scatterer
is Rayleigh scattering by H$_2$, and the dominant absorbers are
alkali metals, particularly K~I (7670 \AA) and Na~I (5894 \AA).
In our approximate models with cloud layers,
the alkali metals are strong, but narrow.
Because the irradiated temperature-pressure profiles cross the condensation boundaries
at relatively low pressure, the strong pressure broadening of the alkali metals does
not occur in our models.

With the choice of a finite cloud layer, an additional
free parameter becomes the location of the cloud base.
For example, in the case considered here the temperature-pressure curve
could cross the Fe and MgSiO$_3$ condensation curves at low $P$, 
where they overlap. In this case a 
Fe-MgSiO$_3$ mix will prevail (see Figure~\ref{fig:TempProfile}).
In contrast, as discussed in \S\ref{sec-TPprofile}, if the cloud
is low in the atmosphere (around 1 bar e.g. Sudarsky et al. 1999) the incoming
radiation will be absorbed by broad
lines of Na~I and K~I before reaching the scattering cloud,
causing zero optical albedo redward of 500~nm.

We have rudimentarily explored finite cloud layers where 
all clouds are 1 pressure scale height above their base
at the equilibrium condensation curve.
The main difference in reflected light curves from our vertically homogenous
assumption is an increase in magnitude.
The reason is that MgSiO$_3$, which has the coolest condensation curve
(shown in Figure~\ref{fig:TempProfile}), would be the top layer;
MgSiO$_3$ is both the most reflective (see $\tilde{\omega}$ in Figure~\ref{fig:g})
and has the largest backscattering probability of the three condensates
considered here.
The light curve shape changes little with
the cloud layer models. In the case of small particles compared to wavelength 
($\overline{r}=0.01~\mu$m for 5500 \AA),
the shape of the phase function is Rayleigh-like for all three
condensates.
In the case for large particles compared to wavelength
($\overline{r}=10~\mu$m for 5500 \AA), $\tilde{\omega}$ of MgSiO$_3$
already dominates
in the three condensate mix for large particles.
However, we caution that much more work needs to be done both
in cloud models and particle size distribution.
We emphasize the difficulty in predicting the reflected light curves
due to the large parameter space of irradiative heating, cloud models,
and particle type and size distribution. Thus these exploratory models
should be considered as a useful interpretative tool rather than a predictive tool.

\subsection{Comparison with Observations}
\subsubsection{Spectral Separation}
Cameron et al. (1999) and Charbonneau et al. (1999) have given the first
observational results for a CEGP atmosphere,
$\tau$~Boo~b. The results are marginally in conflict
(see below); the first
group claims a probable detection with a flux-ratio
of $\epsilon = 1.9 \times 10^{-4}$ and the second group a null result
with upper limit of $\epsilon = 5 \times 10^{-5}$.
Although in this paper we are modeling 51~Peg~b, we can assume to
first order that $\tau$~Boo~b has a similar atmosphere.
However, $\tau$~Boo~b is hotter than 51~Peg~b and may
be heated above the condensation boundary of some grain species.

To first order such observations are extremely useful
in addressing whether or not there are reflective
clouds near the top of the atmosphere.
For example, if the probable Cameron et al. result 
is confirmed, $\tau$ Boo~b
must have a very reflective cloud of particles fairly high
in the planet atmosphere. The reason for this is
the orbital inclination was measured to be 29$^{\circ}$, meaning
only small phase angles are observed during the planet's orbit
and the planet must be very bright to be detectable at all.
If the particles are spherical,
they must also be smaller than the
wavelength of light, because the strongly peaked wavefunctions
of highly reflecting spherical particles (e.g. particles
with $\overline{r}=10~\mu$m with light curve shown in Figure~\ref{fig:lc10})
generate a small light curve amplitude at low inclinations.
Cameron et al. also find a wavelength dependence of the albedo.
If confirmed this will reveal molecular absorbers in the atmosphere,
but not cloud characteristics (which are grey at optical wavelengths).

The Charbonneau et. al result is inclination dependent,
and provides useful constraints if the system is at high orbital inclination.
Their upper limit of $p=0.3$ at high
inclinations excludes atmospheres with extremely 
reflective clouds (such as pure MgSiO$_3$ clouds of small particles.)
Bright models with light curve shapes different from isotropic,
e.g. those shown in Figures~5~and~6 (which have $p>0.3$ for high $i$)
are not excluded, because of the phase function assumption (see below).
Their upper limit is based
on the assumption that the reflected planetary light is an
exact copy of the stellar light for 4668 to 4987 \AA.
For lower inclinations ($i \approx 30^{\circ}$),
Charbonneau et al. give an upper limit of $1 \times 10^{-4}$.
This does not provide a useful model
constraint because at low inclinations only small phases
of the planet are visible, and only a few extreme cases
of highly reflective clouds can be excluded.

Beyond characterizing the very general cloud reflectance
property, the spectral separation observations cannot constrain
the  particle type or size distribution.
One reason is that the observations measure a combination
of the geometric albedo and planet area: $p (R_P/D)^2$.
The planet radius is not known, except for a transiting
planet. Cameron et al. assume a Jupiter-like albedo
of $p=0.55$ and derive $R_P=1.8R_J$; Charbonneau et al. assume
$R_P=1.2R_J$ (based on evolutionary models from Guillot et al. 1996)
and derive an upper limit for $p$.

A second difficulty in using the spectral separation results
to constrain cloud details is the phase function 
assumption that goes into the results.
The results given by both groups are
expressed as a flux ratio or geometric albedo at opposition.
Neither group observed near opposition (``full phase''
for $i=90^{\circ}$), but instead
extrapolated their results based on an assumed phase function.
Cameron et al. used an empirically determined polynomial
approximation to the phase function of Venus (Hilton 1992),
which also resembles that of Jupiter (Hovenier \& Hage 1989).
Charbonneau et al.
used the Lambert sphere phase function, which derives from
isotropic scattering and is approximately valid for Uranus and Venus.
The differences in $p$ from using these different phase function
assumptions is 20\% and this may be one contribution
to the conflicting observational results. For more details
see Cameron et al. (2000) and Charbonneau \& Noyes (2000).
The inherent uncertainty in the albedo measurement from
both the phase function and the radius/albedo degeneracy prevents
any serious constraint on atmosphere models.
Nevertheless, for comparison Table~2 shows the geometric albedos
of our 4 models at 4800~\AA, roughly the center
of both groups' wavelength range.

Results from this paper show that the light curve observations
needed to constrain atmosphere models are those 
at different colors in narrow wavelength bands. The specific
anisotropic scattering properties 
for a given particle size distribution and grain
indicies of refraction will determine the light curve.
The indicies of refraction are both wavelength- and
particle type-dependent, which is why the color dependence
is important. Opacity effects in a narrow
wavelength range are less important because the condensates
are generally grey in the optical.

\subsubsection{Ground-based Photometric Light Curves}
Observations by G. Henry (private communication)
with ground-based automatic photometric telescopes can currently 
reach a precision of near 100~$\mu$mag and could be as precise as 50~$\mu$mag
with a dedicated automatic photometric telescope.
The precision is attainable with observations over
many orbital periods because the phase effect is strictly repeating.
With this limit, reflected light detections of high-orbital inclination systems
should be possible, or at least useful constraints on the models.
Furthermore, these photometric limits should allow confirmation
of the Cameron et al. result. Their result
of $\epsilon = 1.9 \times 10^{-4}$, which using the
polynomial approximation to Venus
(Hilton 1992) for $i = 29^{\circ}$  translates roughly
into a flux-ratio of $9 \times 10^{-5}$ and an amplitude
of $7 \times 10^{-5}$ which corresponds to $\approx 80~\mu$mag.

\subsection{Other EGPs}

\subsubsection{Close-in EGPs}
Because $\epsilon \sim (R_P/D)^2$, the flux
variation of other CEGPs can be estimated using the data in
Table 1, and multiplying the light curves by the
$(D_{51 Peg}/D_{EGP})^2$ ratio. However $\Delta m$ is a flux ratio
and is not affected by the magnitude of the parent star, although
magnitude is observationally important.
This estimate also assumes that the radii of the planets are the same
(cf. Guillot et al. 1996). This estimate, shown for a Lambert sphere in
Figure~\ref{fig:lambert}, shows a variation of a factor of 2
between $D=0.042$~AU and $D=0.059$~AU.

\begin{figure}
{\plotfiddle{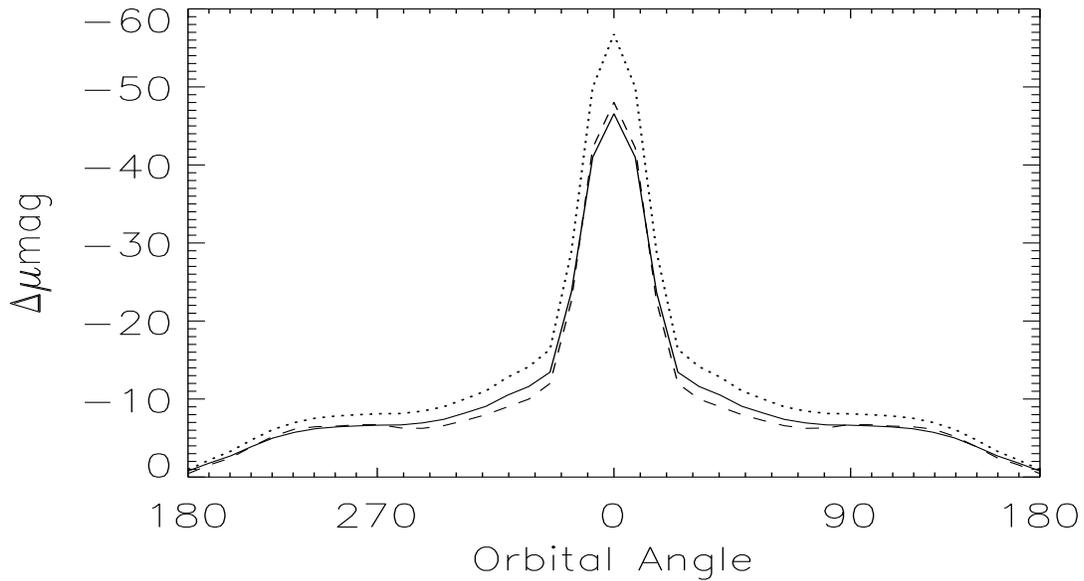}{3.5in}{0}{90}{70}{-85}{-250}}
\caption{Predicted light curves for 51 Peg b (solid curve) and $\tau$ Boo b
(dashed curve) for $i=82^{\circ}$ and particles with $\overline{r}=10~\mu$m. The dotted
curves is for a 51 Peg b-type planet with $D_{\tau Boo}$.}
\label{fig:lcurve51Peg}
\end{figure}

There are additional differences among the different CEGPs that affect
scattering.
One is from density.
For example, 51 Peg b has almost an order of magnitude lower minimum
mass than $\tau$ Boo b for the same planetary radius. A less dense
atmosphere has a longer photon mean free path, which has two effects. One is
more backscattered light. The second, which is minor, is more
unscattered radiation passing through the limb, and less
forward-scattered radiation traveling through the upper atmosphere.
Coincidentally, some of the enhanced scattering effects gained from 51
Peg b's lower surface gravity
atmosphere compared to $\tau$ Boo b are lost with the larger
distance from the parent star. With a lower surface gravity, 51~Peg~b's
atmosphere is less dense (has a lower $P_g$) for the same Rosseland
mean optical depth.
Figure~\ref{fig:lcurve51Peg} shows the light curves for both $\tau$
Boo b and 51 Peg b, for the three condensate mix of
particles with $\overline{r}=10$~$\mu$m, at
$i=82^{\circ}$, which is
not excluded by a transit non-detection.
Also plotted in Figure~\ref{fig:lcurve51Peg} is 51 Peg b's light curve
at $D_{\tau Boo}$.
Because of
its lower surface gravity atmosphere, 51 Peg~b shows effects from more
scattering;
a higher backscattering peak ($ 335^{\circ} < \theta_S < 25^{\circ}$),
and more light
scattered through the upper atmosphere ($\theta_S > 170^{\circ}$).
The difference at $ 25^{\circ} < \theta_S < 335^{\circ}$ is also due
to the different atmosphere densities.
Figure~\ref{fig:lcurve51Peg} shows that the observations would not
be able to constrain the density; away from $\Theta=0^{\circ}$
the differences are very small, and near $\Theta=0^{\circ}$
the amplitude difference is degenerate with change in density, $\tilde{\omega}$,
$R_p$, etc.

A planet with a larger radius than we have assumed,
such as $R_P=1.40 \pm 0.17 R_J$ derived from the transit
of HD~209458~b (Mazeh et al. 2000) would have 
a larger reflected light signal.
The density effects described above would
be a secondary effect.

Discrete cloud layers (not modeled here but
discussed in \S\ref{sec-CloudLayers}) could have an 
effect on the CEGPs.
The bases of different cloud types may be at different
depths, as on Jupiter, because thermodynamical
equilibrium calculations predict condensation curves with different
temperature and pressure dependencies for each
species (see Figure~\ref{fig:TempProfile}).
CEGPs that have a hotter parent star or a smaller $D$
will have different temperature-pressure structures than the
cooler ones; they may be different enough
that different cloud layers are more or less visible.
For example, we find that for
particles with $\overline{r} < 1~\mu$m, $\tau$ Boo b's atmosphere may be heated
enough to have a temperature above that of MgSiO$_3$ condensation.

\subsubsection{EGPs Beyond 0.05 AU}
\label{sec-EGPbeyond}
Because we are investigating planets that could be detectable in reflected
light in the near future, we focus on the CEGPs. However, for
a very rough estimate of EGPs with $D>0.05$~AU, 
the light and polarization curves in this paper can be
scaled by $(D_{51 Peg}/D_{EGP})^2$ (since $\epsilon \sim 1/D^2$).
This rough estimate ignores the
cloud particles, which would be different than those in the
CEGP atmospheres. An EGP such as
$\rho^1$ Cnc, at 0.12~AU from its parent star, 
is 2.35 times as far as 51
Peg b from its parent star.
The maximum amplitude light curve in this paper
(60~$\mu$mag at $i=90^{\circ}$) would be 
only be 11~$\mu$mag at the distance of $\rho^1$ Cnc.
This variation is barely detectable
by the upcoming microsatellite missions. An EGP such as $\rho$ CrB
at  $D=0.23$~AU is 4.5 times farther from its parent
star than 51 Peg b.
The maximum amplitude light curve at this distance is
only 3~$\mu$mag. Such an EGP
will not be photometrically detectable in the foreseeable
future. Thus the CEGPs have the brightest prospects for detection.

\section{Summary and Conclusion}
We have presented photometric light curves and fractional polarization curves
for 51 Peg b for 4 mean particle sizes, and discussed the differences
with color and with other CEGPs. The 
light curves are very sensitive to condensate type and size
distribution, hence observations will be able to distinguish
between extreme scenarios. However, more detailed information
such as the exact size distribution and particle type
will be more difficult to extract.

The temperature-pressure profiles are also
extremely dependent on the condensate assumptions.
We have briefly discussed these, along with the emergent
spectra. In contrast to T dwarfs which
have no observeable clouds, the CEGPs should have clouds closer
to the top of their
atmospheres because irradiation heats the upper atmosphere
to temperatures closer 
to the equilibrium condensation curve (i.e. the cloud base).
The condensates that may not contribute to reflected light
because they are sequestered below the top cloud layer
will still affect the 
temperature-pressure profile by heating the lower
atmosphere.
Thus observations of the light curves which should constrain
the general cloud properties will help distinguish between
atmosphere models.

The light curves may be very different from
sine curves; their shape depends not just on the particle size and type, but
also on $\tilde{\omega}$ and the atmospheric density.
Inclinations other than the narrow angular range possible for
transits are theoretically detectable for the CEGPs.
Many cases of the light curves are
detectable by upcoming space missions, and some
of the largest amplitude cases (e.g.
from pure MgSiO$_3$ particles with $\overline{r}=0.1~\mu$m)
might be detectable from the ground in
the near future. See Table~2, and equation~(\ref{eq:dm}) using
$\epsilon = p(R_P/D)^2$.
Results from this paper show that the light curve observations
that would best constrain atmosphere models are those
at different colors in narrow wavelength bands.

Geometric albedos at 5500 \AA~in this study for a cloud mix of
particles of
MgSiO$_3$, Fe, and Al$_2$O$_3$ range from $p=0.44$ for particles with
$\overline{r}=10~\mu$m that have a strong backscattering peak,
to $p=0.0013$ for particles with $\overline{r}=0.01~\mu$m which include highly
absorbing Fe. Clouds of pure MgSiO$_3$ are much more reflective
for all particle sizes (see Table~2).

The polarization of the CEGP systems is not detectable with current
techniques. Rayleigh scattering polarization peaks at orbital angle
90$^{\circ}$,
but with modulation of the reflected light the fractional polarization
has an asymmetric peak at around 70$^{\circ}$.
Polarization from particles that are large compared to the wavelength will have
more than one peak due to interference effects from different light ray paths
through the particle.

Other CEGP systems in our study give similar light curves to 51 Peg b,
but effects
from distance from the parent star and density are important.
Planets farther than 0.1~AU from their parent stars are too
faint in reflected light to be detected photometrically in the
foreseeable future.

We emphasize that there are many unknowns in the model atmospheres, and
much room for improvement --- mainly more realistic cloud modeling,
heat redistribution by winds, photochemistry, non-spherical particles, and
other types of condensates. These ingredients will result in different
light curves than those shown in this paper.
That the predictions are so varied means observations
should be able to identify the gross cloud characteristics.
In this sense the theory should be seen as an interpretive
rather than a predictive tool.
Observations by the upcoming satellites MOST, COROT, and MONS and
ground-based work will help constrain the CEGP atmosphere models and at best will
reveal the nature of their atmospheres directly.

\bigskip

\acknowledgements{We thank Kenny Wood for illuminating discussions on
the Monte Carlo scattering method, Dave Charbonneau for
many useful discussions, Jens Falkesgaard for many of the
Gibbs Free energy fits. We also thank Mark Marley, Bob Noyes,
Mike Wolff, Greg Henry, and Adam Burrows for useful discussions. 
We thank the referee Tristan Guillot for helpful comments
that improved the paper. SS is supported by
NSF grant PHY-9513835, BAW acknowledges support by NAG5-8587,
and  DDS acknowledges support from the Alfred P. Sloan Foundation.}

\pagebreak
\begin{table}[ht]
The Close-in Extrasolar Giant Planets \\
\begin{tabular}{l c c c c c l}
\tableline
\tableline
 Star Name & Spectral Type &D& $M \sin i$ &P& $T_{\rm eq}(1-A)^{-1/4}$ & Reference \\
 & & (AU) & ($M_J$) & (days) & (K) \\
\tableline
 HD 187123  & G3V  & 0.042 & 0.52  & 3.097 & 1400 & 1 \\
 HD 75289   & G0V  & 0.046 & 0.42  & 3.51  & 1600 & 2 \\
 $\tau$ Boo & F7V  & 0.0462 & 3.87 & 3.3128 & 1600 & 3 \\
 HD 209458  & G0V  & 0.0467 & 0.69 & 3.525 & 1500 & 4\\
 51 Peg     & G2V  & 0.051  & 0.47 & 4.2308 & 1300 & 5 \\
\tableline
\end{tabular}
\tablerefs{(1) Butler et al. 1998; (2) Mayor et al. 1999;
(3) Butler et al. 1997; (4) Mazeh et al. 2000; (5) Mayor \& Queloz 1995}
\end{table}

\begin{table}[hb]
Geometric Albedos \\
\begin{tabular}{l c c c}
\tableline
\tableline
Mean Particle& $\lambda=5500$~\AA & $\lambda=5500$~\AA & $\lambda=4800$~\AA \\
Size  & & MgSiO$_3$ only & \\
\tableline
$0.01 \mu$m  & 0.0013 & 0.18 & 0.0013 \\
$0.1 \mu$m   & 0.18   & 0.69 & 0.14   \\
$1 \mu$m     & 0.41   & 0.50 & 0.36   \\
$10 \mu$m    & 0.44   & 0.55 & 0.4   \\
\tableline
\end{tabular}
\tablecomments{Geometric albedos for the models discussed in this paper
(column 1), for pure MgSiO$_3$ clouds which are highly reflective (column 2),
and for the models in this paper at $\lambda=4800$~\AA~(third column),
which corresponds to the Cameron et al. (1999) and
Charbonneau et al. (1999) observations.}
\end{table}

\pagebreak

\pagebreak

\pagebreak

\end{document}